# Magnetic Coordinate Systems

K.M. Laundal[1]  · A.D. Richmond[2]

**Abstract** Geospace phenomena such as the aurora, plasma motion, ionospheric currents and associated magnetic field disturbances are highly organized by Earth's main magnetic field. This is due to the fact that the charged particles that comprise space plasma can move almost freely along magnetic field lines, but not across them. For this reason it is sensible to present such phenomena relative to Earth's magnetic field. A large variety of magnetic coordinate systems exist, designed for different purposes and regions, ranging from the magnetopause to the ionosphere. In this paper we review the most common magnetic coordinate systems and describe how they are defined, where they are used, and how to convert between them. The definitions are presented based on the spherical harmonic expansion coefficients of the International Geomagnetic Reference Field (IGRF) and, in some of the coordinate systems, the position of the Sun which we show how to calculate from the time and date. The most detailed coordinate systems take the full IGRF into account and define magnetic latitude and longitude such that they are constant along field lines. These coordinate systems, which are useful at ionospheric altitudes, are non-orthogonal. We show how to handle vectors and vector calculus in such coordinates, and discuss how systematic errors may appear if this is not done correctly.

**Keywords** Magnetic coordinates · Ionospheric electrodynamics

## 1 Introduction

The influence of Earth's magnetic field extends to approximately 15 $R_E$ ($R_E$ is the mean Earth radius $\approx$ 6371.009 km) in the hemisphere facing the Sun, and up to several hundred

K.M.L. was supported by the Research Council of Norway/CoE under contract 223252/F50. A.D.R. was supported in part by NSF award AGS-1135446. The National Center for Atmospheric Research is sponsored by the National Science Foundation. The authors thank Nils Olsen, Christopher C. Finlay and Astrid Maute for valuable discussions and proofreading of the manuscript.

✉ K.M. Laundal
  karl.laundal@ift.uib.no

[1] Birkeland Centre for Space Science, University in Bergen, Bergen, Norway

[2] National Center for Atmospheric Research, High Altitude Observatory, Boulder, CO, USA

    🌐 Springer

$R_E$ on the night side. At ionospheric altitudes, the Earth's main field is orders of magnitudes stronger than even the strongest magnetic disturbances created by the interaction with the solar wind. To understand the dynamics of near Earth plasma, electromagnetic field disturbances, and even the neutral atmosphere at high altitudes, it is essential to work in a reference frame which takes the geometry of the Earth's magnetic field into account. Exactly how this should be done differs by the application, region, and desired level of precision.

At large distances from the Earth, the dominating driver is the solar wind flowing roughly radially out from the Sun. The interaction of the solar wind with the magnetosphere in these regions is therefore often best understood in a coordinate system which has the Earth–Sun line along one of the axes. Examples of such coordinate systems include the geocentric solar ecliptic (GSE) and the geocentric solar magnetic (GSM) coordinate systems. The latter also contains the Earth's magnetic dipole field in one of the coordinate planes, making it especially suitable for working with solar wind-magnetosphere interaction processes. Closer to the Earth the relative importance of the Earth's field becomes stronger, so that a more precise alignment with the dipole axis is required. Solar magnetic (SM) coordinates is one such system, having one axis along the dipole axis, and the Earth–Sun line in a coordinate plane.

At ionospheric heights and on ground, the centered dipole (CD) coordinate system is probably the most commonly used magnetic coordinate system. This system, which is most used as spherical coordinates, represents a shift of the poles from the rotational axis to the dipole axis. At these altitudes however, the Earth's field deviates significantly from a centered dipole. To achieve better accuracy with respect to the magnetic field, several alternative coordinate systems exist. One example is the eccentric dipole (ED) coordinate system, which is based on a dipole representation which not necessarily has its origin at the center of the Earth.

Even better accuracy can be achieved by taking the non-dipole features of the Earth's field into account. Corrected geomagnetic (CGM) coordinates (Gustafsson et al. 1992; Baker and Wing 1989; Shepherd 2014, and references therein) and the Magnetic Apex Coordinate systems (Richmond 1995) are based on a more detailed model of the Earth's field, and defined in terms of field line tracing. The improved accuracy comes at the expense of simplicity, as the result is a non-orthogonal coordinate system, in contrast to the aforementioned systems which can be defined in terms of Cartesian base vectors and converted between by means of rotation matrices.

In this review paper, we define all the above mentioned coordinate systems. The definitions of the orthogonal coordinate systems are given such that it is possible to implement coordinate conversion routines using only the most basic inputs: The first few Gauss coefficients in the International Geomagnetic Reference Field (IGRF) model, and time. The paper differs from other reviews on space physics coordinates (Matsushita and Campbell 1967; Russell 1971; Fraser-Smith 1987; Fränz and Harper 2002; Barton and Tarlowski 1991; Hapgood 1992) most notably by the inclusion of the non-orthogonal coordinate systems, which we also explain how to use with vector quantities.

In Sect. 2 we describe the geocentric and geodetic coordinate systems. Section 3 contains the definitions of the orthogonal magnetic coordinate systems, and Sect. 4 the non-orthogonal coordinate systems. Section 5 explains how to handle electrodynamic vector quantities in such systems. Section 6 is about magnetic local time, and the secular variation of magnetic coordinate systems is briefly discussed in Sect. 7. The appendices contain a list of abbreviations (Appendix A), notations (Appendix B), and code to calculate the position of the Sun in geocentric coordinates (Appendix C).

## 2 Geocentric and Geodetic Coordinates

Geocentric spherical coordinates are comprised of the geocentric radial distance $r$, the colatitude $\theta$, and the east longitude $\phi$. A position vector $\mathbf{r}$ can be written in geographic Cartesian coordinates $x, y, z$ as

$$\mathbf{r} = \begin{pmatrix} x \\ y \\ z \end{pmatrix} = \begin{pmatrix} r \sin\theta \cos\phi \\ r \sin\theta \sin\phi \\ r \cos\theta \end{pmatrix} \qquad (1)$$

so that the spherical coordinates can be expressed in terms of $x, y, z$ as:

$$\begin{aligned} r &= \sqrt{x^2 + y^2 + z^2} \\ \theta &= \cos^{-1}\left(\frac{z}{\sqrt{x^2 + y^2 + z^2}}\right) \\ \phi &= \mathrm{atan2}(y, x). \end{aligned} \qquad (2)$$

The term geocentric can refer to any Cartesian coordinate system which has its origin at the center of the Earth. In this paper (unless explicitly followed by "solar magnetic" or "solar ecliptic") we take geocentric (abbreviation GEO) coordinates to mean the coordinate system which has its $z$ axis ($\hat{\mathbf{z}}_{geo}$) along the rotational axis, the $x$ axis ($\hat{\mathbf{x}}_{geo}$) pointing at the Greenwich meridian and the equator, and the $y$ axis ($\hat{\mathbf{y}}_{geo}$) completing a right-handed system. The geocentric coordinates at $\mathbf{r}$ (1), can be written in terms of these base vectors as

$$\mathbf{r} = \begin{pmatrix} x \\ y \\ z \end{pmatrix} = \begin{pmatrix} \hat{\mathbf{x}}_{geo} \cdot \mathbf{r} \\ \hat{\mathbf{y}}_{geo} \cdot \mathbf{r} \\ \hat{\mathbf{z}}_{geo} \cdot \mathbf{r} \end{pmatrix}. \qquad (3)$$

A local Cartesian coordinate system can also be constructed at a point $(\theta, \phi)$ on $r = R_E$, such that the axes point along meridians towards the North Pole, eastward along circles of latitudes, and upward. We call such a coordinate system ENU (east-north-up), in contrast to the ECEF (Earth-centered Earth-fixed) system described above. The ENU base vectors at $(\theta, \phi)$ are (in geocentric coordinates):

$$\hat{\mathbf{e}} = \begin{pmatrix} -\sin\phi \\ \cos\phi \\ 0 \end{pmatrix}, \qquad \hat{\mathbf{n}} = \begin{pmatrix} -\cos\theta \cos\phi \\ -\cos\theta \sin\phi \\ \sin\theta \end{pmatrix}, \qquad \hat{\mathbf{u}} = \begin{pmatrix} \sin\theta \cos\phi \\ \sin\theta \sin\phi \\ \cos\theta \end{pmatrix}. \qquad (4)$$

If a position vector is given in ECEF coordinates, ENU components can be found by multiplication with a rotation matrix. The rotation matrix for conversion from ECEF to ENU has the transposed ENU base vectors, expressed in geocentric coordinates (4) as rows. We denote this matrix by $\underset{ecef \to enu}{\mathbf{R}}$:

$$[\mathbf{r}]_{enu} = \underset{ecef \to enu}{\mathbf{R}} [\mathbf{r}]_{ecef} = \begin{pmatrix} -\sin\phi & \cos\phi & 0 \\ -\cos\theta \cos\phi & -\cos\theta \sin\phi & \sin\theta \\ \sin\theta \cos\phi & \sin\theta \sin\phi & \cos\theta \end{pmatrix} [\mathbf{r}]_{ecef} \qquad (5)$$

where the bracket subscripts denote the coordinate system in which the vector components are represented ($\mathbf{r}$ is a position vector without reference to a coordinate system, while $[\mathbf{r}]_{ecef}$ explicitly refers to the components of **r** in ECEF coordinates). Since the base vectors are orthonormal, the matrix for the inverse operation is given by the transpose:

$$\underset{enu \rightarrow ecef}{\mathbf{R}} = \underset{ecef \rightarrow enu}{\mathbf{R}^\top}. \tag{6}$$

The Earth is slightly oblate, and **r** is not precisely normal to its surface except at the poles and equator. The World Geodetic System 1984 (WGS84) datum defines a reference Earth surface as an ellipsoid with equatorial radius $R_{eq} = 6,378,137.0$ m and a reciprocal of flattening $1/f = 298.257223563$, giving a polar radius of $R_p = 6,356,752.3$ m. Geodetic (or geographic) coordinates are defined with respect to this surface. Altitude $h$ is the distance of a point from the surface. The gradient of $h$ gives a unit vertical (upward) vector $\hat{\mathbf{k}}$. Geodetic longitude is identical with geocentric longitude. Geodetic colatitude is the angle between $\hat{\mathbf{k}}$ and the Earth's axis, and differs slightly from geocentric colatitude except at the poles and equator. Geodetic latitude, $\lambda_{gd}$, is 90° minus geodetic colatitude.

The geocentric Cartesian coordinates at a point defined in terms of geodetic longitude, latitude and height, $(\phi, \lambda_{gd}, h)$ can be calculated as

$$\begin{aligned} x &= (\rho + h) \cos \lambda_{gd} \cos \phi \\ y &= (\rho + h) \cos \lambda_{gd} \sin \phi \\ z &= \left(\rho + h - e^2 \rho\right) \sin \lambda_{gd} \end{aligned} \tag{7}$$

where

$$\rho = R_{eq}\left(1 - e^2 \sin^2 \lambda_{gd}\right)^{-1/2} \tag{8}$$

and $e$ is the eccentricity of the ellipsoid, given by

$$e^2 = 2f - f^2. \tag{9}$$

The inverse transformation, from geocentric to geodetic coordinates, is less straightforward, and a large number of different algorithms exist (see e.g., Zhu 1994, and references therein).

Note that the ENU directions in (4) refer to geocentric coordinates, so that "north" is northward on a spherical Earth, and "up" is radially with respect to the center of the Earth. ENU commonly refers to the corresponding geodetic directions, in which case an additional rotation must be applied, about the east direction, to make the northward vector tangent, and the upward vector (which we then call $\hat{\mathbf{k}}$) normal to the ellipsoid.

Geocentric coordinates are used throughout Sect. 3. Geodetic coordinates are used in Sect. 4.

## 3 Orthogonal Magnetic Coordinate Systems

In this section we define the most common orthogonal magnetic coordinate systems in terms of their origin and Cartesian base vectors represented in geocentric coordinates. When these coordinate vectors are known, it is straightforward to construct matrices to rotate between coordinate systems, and to calculate corresponding spherical coordinates by use of (2). The information given here should be sufficient to write computer software for coordinate transformations between all the orthogonal magnetic coordinate systems and geocentric coordinates.

The definitions of all the orthogonal coordinate systems involve the Earth's dipole axis. The dipole model which is normally used is that of the IGRF, which is also the standard magnetic field model used in the nonorthogonal coordinate systems. The IGRF (Thébault et al. 2015), represents the Earth's magnetic field in terms of a series of spherical harmonic functions:

$$\mathbf{B}(r,\theta,\phi,t) = -R\nabla \sum_{n=1}^{N_{\max}} \left(\frac{R}{r}\right)^{n+1} \sum_{m=0}^{n} \left[g_n^m(t)\cos m\phi + h_n^m(t)\sin m\phi\right] P_n^m(\cos\theta), \quad (10)$$

where $r$ is geocentric radius, $\theta$ is geocentric colatitude, $\phi$ is longitude, and $t$ time. $R = 6371.2$ km is a reference radius used in geomagnetic modeling, which is close to the mean Earth radius, $R_E$, calculated as $(2R_{eq} + R_p)/3 \approx 6371.009$ km (using the WGS84 values given in the previous section). $P_n^m$ is the Schmidt seminormalized associated Legendre function of order $m$ and degree $n$, and $g_n^m$ and $h_n^m$ are the Gauss coefficients of the internal field. Since the Earth's field changes, the Gauss coefficients are time-dependent, and the IGRF is updated every five years. The latest version, IGRF-12 (Thébault et al. 2015), is used here. For periods in between each model update, the coefficients should be found by linear interpolation between the closest models.

The truncation level $N_{\max}$ is 13 for the latest IGRF models, so that the sum contains 195 Gauss coefficients. Prior to 2000, $N_{\max} = 10$ (120 coefficients). Only the three first coefficients ($n = 1$) are needed to define the dipole component of the field. A unit vector which is anti-parallel to the dipole axis (the magnetic moment) can be expressed as

$$\hat{\mathbf{m}} = -\frac{1}{B_0} \begin{pmatrix} g_1^1 \\ h_1^1 \\ g_1^0 \end{pmatrix} \quad (11)$$

where

$$B_0 = \sqrt{\left(g_1^0\right)^2 + \left(g_1^1\right)^2 + \left(h_1^1\right)^2} \quad (12)$$

is a reference magnetic field, or reduced moment (see e.g., Fraser-Smith 1987, and references therein). The Earth's field is such that the dipole axis points roughly southward, so that the dipole North Pole is really in the Southern Hemisphere (SH). However convention dictates that the axis of the geomagnetic dipole is positive northward, hence the negative sign in the definition of $\hat{\mathbf{m}}$. The intersection of this vector with a unit sphere gives the positions of the centered dipole poles. Writing $\hat{\mathbf{m}}$ in terms of the centered dipole pole coordinates in the Northern Hemisphere (NH), we get

$$\hat{\mathbf{m}} = \begin{pmatrix} \sin\Theta_N \cos\Phi_N \\ \sin\Theta_N \sin\Phi_N \\ \cos\Theta_N \end{pmatrix}, \quad (13)$$

where the geocentric colatitude and longitude of the pole are represented as capital Greek letters. They are related to the Gauss coefficients by:

$$\begin{aligned}\Theta_N &= \cos^{-1}\left(-\frac{g_1^0}{B_0}\right) \\ \Phi_N &= \operatorname{atan2}\left(h_1^1, g_1^1\right).\end{aligned} \quad (14)$$

In the SH the location of the dipole pole is $\Theta_S = 180° - \Theta_N$, $\Phi_S = \Phi_N + 180°$. With IGRF-12 values for 2015, the pole locations are at $\Theta_N = 9.69°$, $\Phi_N = -72.63°$, and $\Theta_S = 170.31°$, $\Phi_S = 107.37°$.

The vector $\hat{\mathbf{m}}$ is needed in the definitions of all the orthogonal magnetic coordinate systems described below. Some of the coordinate systems also depend on the direction of the Sun in the geocentric system. A unit vector, $\hat{\mathbf{s}}$, pointing in this direction can be expressed in the same way as the components of $\hat{\mathbf{m}}$ in (13), using the geocentric colatitude and longitude of $\hat{\mathbf{s}}$ instead of the pole location. These values of geocentric colatitude and longitude are also the values of geodetic colatitude and longitude of the point on Earth where the Sun is in zenith (the subsolar point), since $\hat{\mathbf{k}}$ at the subsolar point is essentially parallel to $\hat{\mathbf{s}}$, even if the line between the Earth center and the Sun does not precisely pass through the subsolar point owing to the Earth's oblateness. The subsolar point depends only on the time and date. Python code which can be used to calculate its coordinates is given in Appendix C.

The dipole tilt angle, $\psi$, describes the orientation of the dipole axis with respect to the Earth–Sun line. It is defined to be 0° when they are at right angles, and positive when the centered dipole pole in the north is tilted towards the Sun. When $\hat{\mathbf{s}}$ and $\hat{\mathbf{m}}$ are known, it can be calculated as

$$\psi = \sin^{-1}(\hat{\mathbf{s}} \cdot \hat{\mathbf{m}}). \tag{15}$$

## 3.1 Centered Dipole (CD) Coordinates

The centered dipole coordinate system is arguably the most commonly used magnetic coordinate system. It is often called geomagnetic, geomagnetic dipole, or simply magnetic coordinates, and abbreviated MAG (e.g., Russell 1971). We use the term centered dipole, and abbreviation CD, which is more descriptive in terms of its definition, and to distinguish between this and the other magnetic coordinate systems. Earth-fixed magnetic coordinate systems such as centered dipole coordinates are often used for magnetically organized phenomena close to ground, for instance to represent ground magnetic field perturbations or to organize measurements of geomagnetic disturbances.

Centered dipole coordinates are defined so that the Cartesian $z$ axis aligns with the dipole axis, pointing towards the centered dipole pole in the NH (i.e. in the direction of $\hat{\mathbf{m}}$). The $y$ axis is perpendicular to the plane containing the dipole axis and the rotation axis of the Earth ($\hat{\mathbf{z}}_{geo}$). The $x$ axis completes a right-handed system. Mathematically the base vectors are:

$$\begin{aligned} \hat{\mathbf{z}}_{cd} &= \hat{\mathbf{m}} \\ \hat{\mathbf{y}}_{cd} &= \frac{\hat{\mathbf{z}}_{geo} \times \hat{\mathbf{z}}_{cd}}{\|\hat{\mathbf{z}}_{geo} \times \hat{\mathbf{z}}_{cd}\|} \\ \hat{\mathbf{x}}_{cd} &= \hat{\mathbf{y}}_{cd} \times \hat{\mathbf{z}}_{cd}. \end{aligned} \tag{16}$$

In the hypothetical case that the dipole and rotational axes align, the denominator in the equation for $\hat{\mathbf{y}}_{cd}$ is zero. In that case the CD coordinates can be set equal to geocentric coordinates. Centered dipole coordinates are most often referred to in spherical coordinates (see (2)). A map of centered dipole coordinates is shown in Fig. 1.

If a position vector is given in geocentric coordinates, the centered dipole components can be found by multiplication with a rotation matrix, as in (5). The rotation matrix for conversion from geocentric to centered dipole coordinates has the transposed CD base vectors (16), expressed in geocentric coordinates, as rows. We denote this matrix by $\underset{geo\to cd}{\mathbf{R}}$:

$$[\mathbf{r}]_{cd} = \underset{geo\to cd}{\mathbf{R}} [\mathbf{r}]_{geo} = \begin{pmatrix} [\hat{\mathbf{x}}_{cd}^\top]_{geo} \\ [\hat{\mathbf{y}}_{cd}^\top]_{geo} \\ [\hat{\mathbf{z}}_{cd}^\top]_{geo} \end{pmatrix} [\mathbf{r}]_{geo} \qquad (17)$$

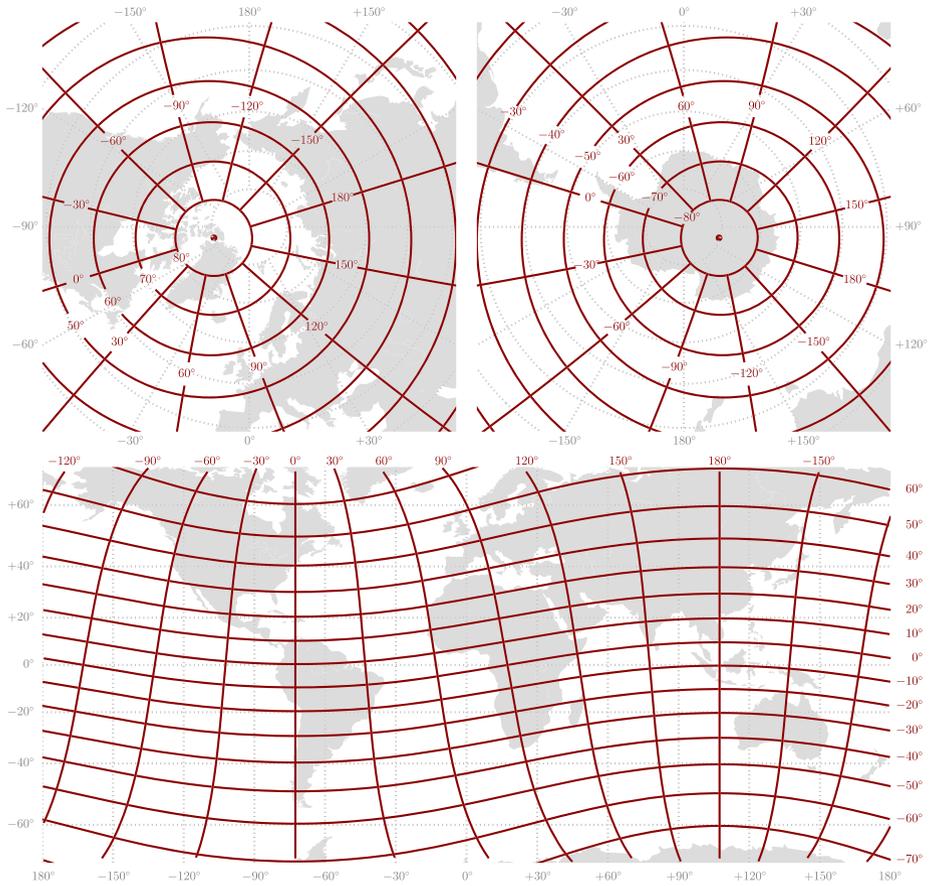

**Fig. 1** Map of centered dipole coordinates. Stereographic projection is used for the polar maps, and Miller cylindrical projection for the world map

Since the base vectors are orthonormal, the matrix for the inverse operation is given by the transpose:

$$\underset{cd\to geo}{\mathbf{R}} = \underset{geo\to cd}{\mathbf{R}}^\top. \qquad (18)$$

An ENU basis in centered dipole coordinates ($\hat{\mathbf{e}}_{cd}$ pointing eastward along CD circles of latitude and $\hat{\mathbf{n}}_{cd}$ pointing along CD meridians) can be found by applying $\underset{geo\to cd}{\mathbf{R}}$ to $\hat{\mathbf{e}}, \hat{\mathbf{n}}$, and $\hat{\mathbf{u}}$ as defined in (4). A matrix whose rows are the resulting vectors is the change of coordinate matrix from geocentric ECEF to centered dipole ENU coordinates.

### 3.2 Local Magnetic Coordinates

Magnetic field measurements are often represented by so-called magnetic elements, three of which are $H$, $D$ and $Z$, where $H$ is the horizontal field strength, $D$ is the declination angle (angle with geodetic north, positive eastward) and $Z$ is the downward component of the field. Magnetic perturbations to the main field are often presented with an $H$ component with a somewhat different meaning; the component of the perturbation in the direction of the main field, i.e. in the compass direction. This direction is often empirically determined using geomagnetic quiet days or other techniques to determine a baseline magnetic field. The other components of such a local magnetic coordinate system are often local magnetic east (90° clockwise of the $H$ direction, seen from above) and up.

If $\hat{\mathbf{b}}$ is a unit vector in the direction of **B** and $\hat{\mathbf{k}}$ is a unit upward vector, then unit vectors in the local magnetic east ($\hat{\mathbf{e}}_m$) and north ($\hat{\mathbf{n}}_m$) directions are

$$\hat{\mathbf{e}}_m = \frac{\hat{\mathbf{b}} \times \hat{\mathbf{k}}}{\|\hat{\mathbf{b}} \times \hat{\mathbf{k}}\|} \tag{19}$$

$$\hat{\mathbf{n}}_m = \hat{\mathbf{k}} \times \hat{\mathbf{e}}_m. \tag{20}$$

A unit vector perpendicular to **B** in the local magnetic-meridional plane, positive upward/poleward, is

$$\hat{\mathbf{p}} = \hat{\mathbf{e}}_m \times \hat{\mathbf{b}}. \tag{21}$$

The local magnetic field is also used to define the dip latitude:

$$\lambda_{dip} = \tan^{-1}\left(\frac{1}{2}\tan I\right) = \tan^{-1}\left(\frac{Z}{2H}\right), \tag{22}$$

where $I$ is the downward dip angle, determined from the horizontal magnetic field strength $H$ and the downward component $Z$. If the Earth was perfectly spherical and the magnetic field a perfect dipole, the dip latitude would be equal to the CD latitude (e.g., Emmert et al. 2010). In reality however, dip latitude is significantly different from CD latitude. Contours of dip latitude, calculated using the IGRF-12, are shown in Fig. 2, which also shows the location of the CD poles as blue dots. The dip poles are located where the horizontal field is zero. These are the magnetic poles at which one would arrive by following a compass.

Local coordinates are easy to determine from local measurements of the field, but are not generally conjugate to local coordinates at other places along a field line, like the opposite hemisphere. Interhemispheric differences between dip latitudes for high-latitude field lines can become quite large, and variations of dip latitude along a meridian are not always monotonic, as near South Africa in Fig. 2.

### 3.3 Eccentric Dipole (ED) Coordinates

The centered dipole upon which centered dipole coordinates are based depends on the three first coefficients in a spherical harmonic representation of the Earth's magnetic field. These coefficients are enough to account for about 95% of the Earth's magnetic field (Lowes 1994). It is possible to have a dipole description of a larger fraction of the field by treating its origin

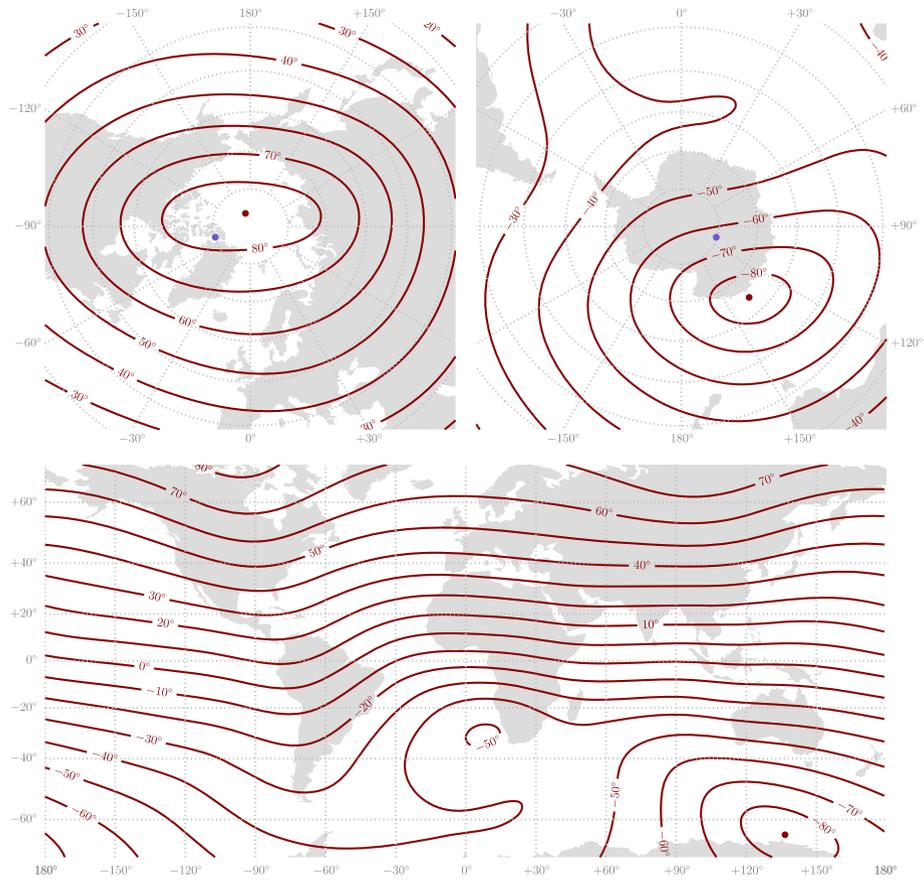

**Fig. 2** Contours of dip latitude, calculated using (22) and IGRF-12 2015 coefficients. The *blue dots* mark the CD poles. The projections are similar to Fig. 1

and orientation as free parameters. However, according to Lowes (1994), who described five different techniques to estimate such a dipole, "there is no optimum eccentric dipole with which to represent the geomagnetic field approximately".

The most used eccentric dipole, introduced by Schmidt (1934), has the same magnitude and orientation as that of the centered dipole, but its position $C$ is chosen such as to minimize the $n = 2$ terms in (10) as observed from $C$. The origin of this eccentric dipole therefore depends on the first eight Gauss coefficients. Here we present formulas which can be used to calculate its position, and refer to Fraser-Smith (1987, and references therein) for more details about their derivation. The origin of the Schmidt eccentric dipole is located at (in Cartesian geocentric coordinates):

$$\Delta x = \eta R_E \quad (23)$$

$$\Delta y = \zeta R_E \quad (24)$$

$$\Delta z = \xi R_E \quad (25)$$

where

$$\eta = (L_1 - g_1^1 E)/3B_0^2 \tag{26}$$

$$\zeta = (L_2 - h_1^1 E)/3B_0^2 \tag{27}$$

$$\xi = (L_0 - g_1^0 E)/3B_0^2 \tag{28}$$

where

$$L_0 = 2g_1^0 g_2^0 + \sqrt{3}(g_1^1 g_2^1 + h_1^1 h_2^1) \tag{29}$$

$$L_1 = -g_1^1 g_2^0 + \sqrt{3}(g_1^0 g_2^1 + g_1^1 g_2^2 + h_1^1 h_2^2) \tag{30}$$

$$L_2 = -h_1^1 g_2^0 + \sqrt{3}(g_1^0 h_2^1 - h_1^1 g_2^2 + g_1^1 h_2^2) \tag{31}$$

$$E = (L_0 g_1^0 + L_1 g_1^1 + L_2 h_1^1)/4B_0^2 \tag{32}$$

and $B_0$ is given by (12).

With the IGRF-12 coefficients for 2015, the shift of the eccentric dipole away from the center of the Earth is

$$\delta = R_E \sqrt{\eta^2 + \zeta^2 + \xi^2} \approx 576.8 \text{ km}. \tag{33}$$

Since 1950, the eccentric dipole (as defined by Schmidt) has moved away from the center of the Earth at an almost constant speed of $\approx 2.5$ km/year.

The base vectors in eccentric coordinates are the same as for centered dipole coordinates (16), since the orientation of the dipole axis is kept the same. That means that coordinate transformations to ED coordinates amounts to a translation from the CD system by a vector **t**, whose geocentric components are $(\Delta x, \Delta y, \Delta z)$, given in (23)–(25). The CD components of **t** are $\Delta x_{cd} = \mathbf{t} \cdot \hat{\mathbf{x}}_{cd}$, $\Delta y_{cd} = \mathbf{t} \cdot \hat{\mathbf{y}}_{cd}$ and $\Delta z_{cd} = \mathbf{t} \cdot \hat{\mathbf{z}}_{cd}$.

Now the relationship between CD and ED components of a position vector is given by

$$\begin{pmatrix} x_{ed} \\ y_{ed} \\ z_{ed} \end{pmatrix} = \begin{pmatrix} x_{cd} - \Delta x_{cd} \\ y_{cd} - \Delta y_{cd} \\ z_{cd} - \Delta z_{cd} \end{pmatrix}. \tag{34}$$

When the Cartesian ED components of a position vector are known, the corresponding spherical coordinates can be found in the standard way (2). Note that the radius $r_{ed}$ in general will not be equal to $r_{cd}$, since the origin has been shifted away from the center of the Earth. Because of the simplicity of converting between ED and CD coordinates, the easiest way to convert between geocentric and ED coordinates is to first convert to CD coordinates.

Sometimes a point is specified by a mixture of CD and ED coordinates. The ED North Pole for instance is located at $\theta_{ed} = 0$ and $r_{cd} = R_E$ (and $\phi_{ed}$ arbitrary). The same problem applies when an ED coordinate grid is defined on the surface of the Earth. To find the CD coordinates in this case, (34) can be used to find a quadratic equation for $r_{ed}$ (which is unknown in this case):

$$\begin{aligned} R_E^2 = &(r_{ed} \sin\theta_{ed} \cos\phi_{ed} + \Delta x_{cd})^2 \\ &+ (r_{ed} \sin\theta_{ed} \sin\phi_{ed} + \Delta y_{cd})^2 \\ &+ (r_{ed} \cos\theta_{ed} + \Delta z_{cd})^2. \end{aligned} \tag{35}$$

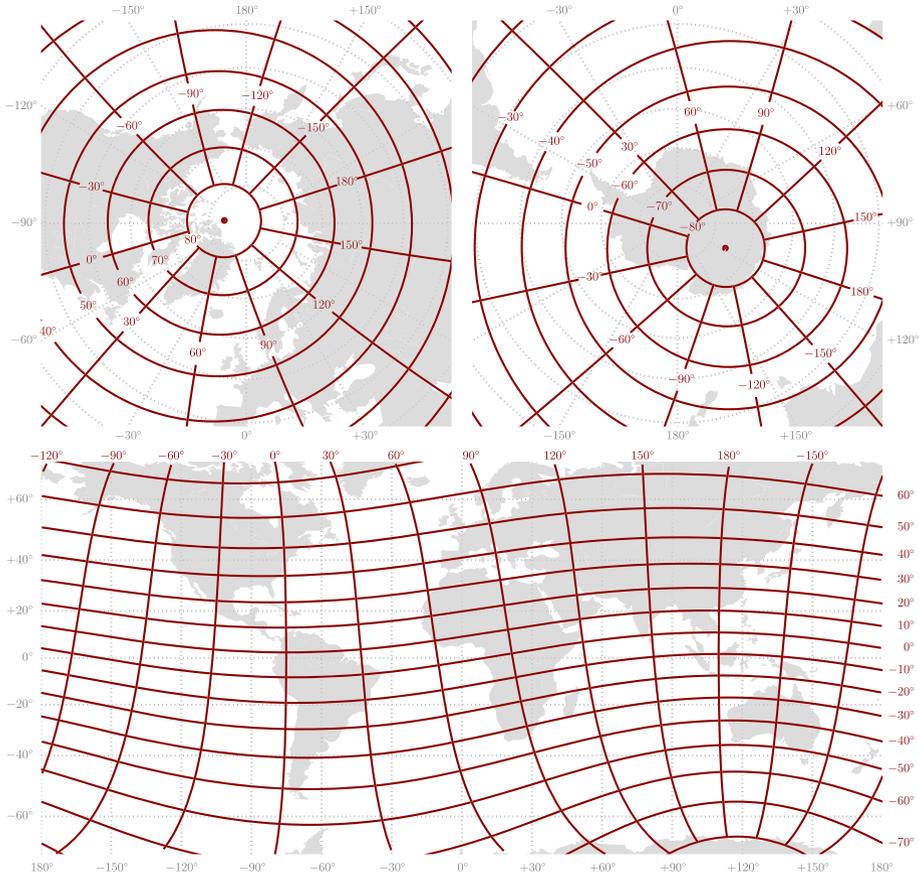

**Fig. 3** Maps of eccentric dipole coordinates at 1 $R_E$. The projections are similar to Fig. 1

Solving this equation for $r_{ed}$ (choosing the positive root), all the Cartesian ED components will be known. The centered dipole components can then be found from (34).

Using this technique, and IGRF-12 coefficients for 2015, we find that the eccentric dipole pole in the NH is at 5.86° colatitude and −97.78° longitude (geocentric coordinates). In the SH it is at 165.72° colatitude and 117.51° longitude. Figure 3 shows maps of ED coordinates at 1 $R_E$.

Fraser-Smith (1987) presents direct relationships between the angular coordinates in centered and eccentric dipole coordinates, based on work by Cole (1963). He also presents expressions for the eccentric dipole pole positions in terms of geocentric coordinates. Apart from an approximation in his expression for pole positions, his formulas give the same results as does the method described above, but computer implementation is more involved, since it requires extra checks to find the correct quadrant for the angles in some cases. This is avoided here if the atan2-function is used.

### 3.4 Geocentric Solar Magnetic (GSM) Coordinates

Geocentric solar magnetic (GSM) coordinates are frequently used at high altitudes; in the solar wind, the magnetopause region and in the magnetotail; regions strongly controlled by the solar wind and the interplanetary magnetic field. The $x$ axis in GSM coordinates points from the center of the Earth (the origin) to the Sun. The $y$ axis is perpendicular to both the magnetic dipole axis and the Earth–Sun line, and positive towards dusk. The $z$ axis, which consequently is in the plane containing both the Earth–Sun line and the dipole axis, completes a right-handed system. The GSM base vectors are:

$$\hat{\mathbf{x}}_{gsm} = \hat{\mathbf{s}}$$
$$\hat{\mathbf{y}}_{gsm} = \frac{\hat{\mathbf{m}} \times \hat{\mathbf{x}}_{gsm}}{\|\hat{\mathbf{m}} \times \hat{\mathbf{x}}_{gsm}\|} \quad (36)$$
$$\hat{\mathbf{z}}_{gsm} = \hat{\mathbf{x}}_{gsm} \times \hat{\mathbf{y}}_{gsm}.$$

GSM coordinates have the same $x$ axis as geocentric solar ecliptic (GSE) coordinates. The $y$- and $z$ axes are defined differently however. In GSE coordinates, which is not a magnetic coordinate system, the $y$ axis is in the ecliptic plane, and the $z$ axis completes the right-handed system (see e.g., Hapgood 1992, for formulas to calculate the GSE axes). Thus in GSE coordinates the dipole axis moves in three dimensions, while in GSM coordinates it moves in only two dimensions, since it is always contained in the $xz$ plane.

Coordinate transformations from GEO to GSM can be done in the same way as in (5), only that the rows of $\underset{geo \to gsm}{\mathbf{R}}$ are the base vectors in (36). A change of coordinate matrix from CD to GSM can be constructed by combining the two matrices:

$$\underset{cd \to gsm}{\mathbf{R}} = \underset{geo \to gsm}{\mathbf{R}} \, \underset{geo \to cd}{\mathbf{R}^\top} \quad (37)$$

### 3.5 Solar Magnetic (SM) Coordinates

Solar magnetic coordinates are much used in regions which are more strongly dominated by the Earth's magnetic field, such as in the inner magnetosphere. Like CD coordinates, the $z$ axis in SM coordinates is the dipole axis. The other axes are defined such that the Sun–Earth line is contained in the $xz$ plane. Thus the difference from CD coordinates is a rotation about the $z$ axis. The $y$ axis, being perpendicular to both the dipole axis and the Sun–Earth line, is the same as in GSM coordinates. The difference between SM and GSM coordinates is thus a rotation about the $y$ axis by the dipole tilt angle (15). In SM coordinates, the dipole axis is static, while the Sun–Earth line moves in the $xz$ plane, in contrast to GSM coordinates, where the Sun–Earth line is static and the dipole axis moves.

The SM base vectors are:

$$\hat{\mathbf{z}}_{sm} = \hat{\mathbf{m}}$$
$$\hat{\mathbf{y}}_{sm} = \frac{\hat{\mathbf{z}}_{sm} \times \hat{\mathbf{s}}}{\|\hat{\mathbf{z}}_{sm} \times \hat{\mathbf{s}}\|} \quad (38)$$
$$\hat{\mathbf{x}}_{sm} = \hat{\mathbf{y}}_{sm} \times \hat{\mathbf{z}}_{sm}$$

Change of coordinate matrices can be constructed in the same way as in (5), with the SM base vectors in place of the CD base vectors. Rotation to SM coordinates from GSM or CD coordinates can be achieved by combining change of coordinate matrices, as in (37). For GSM to SM the change of coordinate matrix can therefore be written:

$$\underset{gsm \to sm}{\mathbf{R}} = \underset{geo \to sm}{\mathbf{R}} \, \underset{geo \to gsm}{\mathbf{R}^\top}. \quad (39)$$

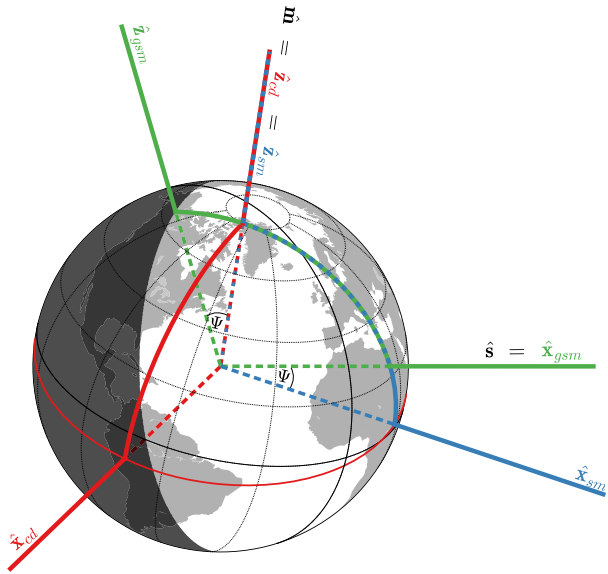

**Fig. 4** Illustration of the *x* and *z* axes of the CD (*red*), SM (*blue*), and GSM (*green*) coordinate systems. The northern half of the prime meridians are also shown. When the axes or meridians of two systems coincide, they are shown in alternating colors. The SM and GSM $xz$ planes are equal, and are spanned by the vectors $\hat{\mathbf{m}}$ (along the dipole axis) and $\hat{\mathbf{s}}$ (pointing to the Sun). Their *x* and *z* axes differ by the dipole tilt angle $\psi$ (15). The GSM *x* axis is along $\hat{\mathbf{s}}$, and the *z* axis intersects the sunlight terminator (the dark side of the Earth is *shaded*). The SM *z* axis is along $\hat{\mathbf{m}}$, and the *x* axis intersects the CD equator, which is also the SM equator (shown by a *thin red great circle*). The CD *z* axis is along $\hat{\mathbf{m}}$, and the *x* axis is in the plane containing $\hat{\mathbf{m}}$ and the Earth's rotational axis. CD coordinates only depend on $\hat{\mathbf{m}}$, and therefore change very slowly with respect to the Earth's surface. SM and GSM coordinates depend also on $\hat{\mathbf{s}}$, which changes with the subsolar point. Orthographic projection is used in this plot

The spherical SM coordinates are similar to spherical CD coordinates apart from a different definition of longitude. The 0° longitude in SM coordinates will contain the subsolar point, and thus represents the centered dipole noon meridian (see Sect. 6 about magnetic local time).

Figure 4 summarizes the relationships between CD, SM, and GSM coordinates. It shows the *x* and *z* axes, and the 90° segment of the great circles (meridians) connecting their intersection with the surface.

## 4 Nonorthogonal Magnetic Coordinate Systems

The magnetic coordinate systems defined in the previous section take into account only the first few terms in the spherical harmonic representation of the Earth's magnetic field (10). Since each term is proportional to $1/r^{n+2}$, the dipole field ($n = 1$) dominates at large distances. Taking into account higher degree terms at high altitudes is therefore often inexpedient. At ionospheric altitudes however, the higher degree terms can be significant. In this section we describe some of the coordinate systems which are defined using the IGRF at full resolution (to degree 13 in 2015). They have the common feature that they are nonorthogonal, and that they would reduce to CD coordinates if $g_n^m$ and $h_n^m$ in (10) were zero for $n, m > 1$ (i.e. a pure dipole field), and the Earth were spherical. The inclusion of higher order terms means that the coordinate systems described here can be used to construct models and organize data with a higher precision with respect to the Earth's magnetic field than with e.g. centered dipole coordinates.

### 4.1 Magnetic Apex Coordinates

The Magnetic Apex coordinate systems (VanZandt et al. 1972; Richmond 1995) are calculated by tracing along magnetic field lines of the IGRF, and defined in terms of the highest point above the Earth, the apex, taking the ellipsoidal shape of the Earth into account. The apex longitude is the centered dipole longitude of the apex. The latitude is determined by the apex height, $h_A$, by mapping back to a certain height along a dipole field line. In the original definition (VanZandt et al. 1972), the Magnetic Apex latitude is

$$\lambda_a = \pm \cos^{-1} \sqrt{\frac{R_{eq}}{R_{eq} + h_A}} \tag{40}$$

where $R_{eq}$ is the equatorial radius of the ellipsoid. The sign of the latitude is positive (negative) north (south) of the dip equator.

Two types of Magnetic Apex coordinate systems were later defined by Richmond (1995), which essentially differ by the height to which the dipole mapping is done: In Quasi-Dipole (QD) coordinates the latitude is defined by mapping to the geodetic height of the point in question, $h$. In Modified Apex coordinates, the latitude is defined by mapping to a reference height $h_R$. Thus the QD latitude is defined as:

$$\lambda_{qd} = \pm \cos^{-1} \sqrt{\frac{R_E + h}{R_E + h_A}}, \tag{41}$$

and the Modified Apex (MA) latitude is defined as

$$\lambda_{ma} = \pm \cos^{-1} \sqrt{\frac{R_E + h_R}{R_E + h_A}}. \tag{42}$$

For both QD and MA coordinates the sign is positive in the Northern magnetic hemisphere and negative for the Southern. It can be seen that Magnetic Apex and Modified Apex coordinates have the property that any point on the same IGRF field line will have the same latitude. The QD latitude, however, depends on the height of that point. Therefore, the QD latitude varies along a field line. Equations (41) and (42) can be used to find an equation relating QD and MA latitudes:

$$\cos \lambda_{qd} = \sqrt{\frac{R_E + h}{R_E + h_R}} \cos \lambda_{ma}. \tag{43}$$

The QD longitude is constant along a field line, and equal to the Modified Apex longitude.

QD coordinates are useful for magnetically organized phenomena that have a specific height distribution, such as ionospheric currents, which are confined to the conducting layer of the ionosphere. The three coordinates in QD coordinates are $\phi_{qd}, \lambda_{qd}$ and $h$, whereas in Modified Apex coordinates the third coordinate is the magnetic potential $V$, which relates

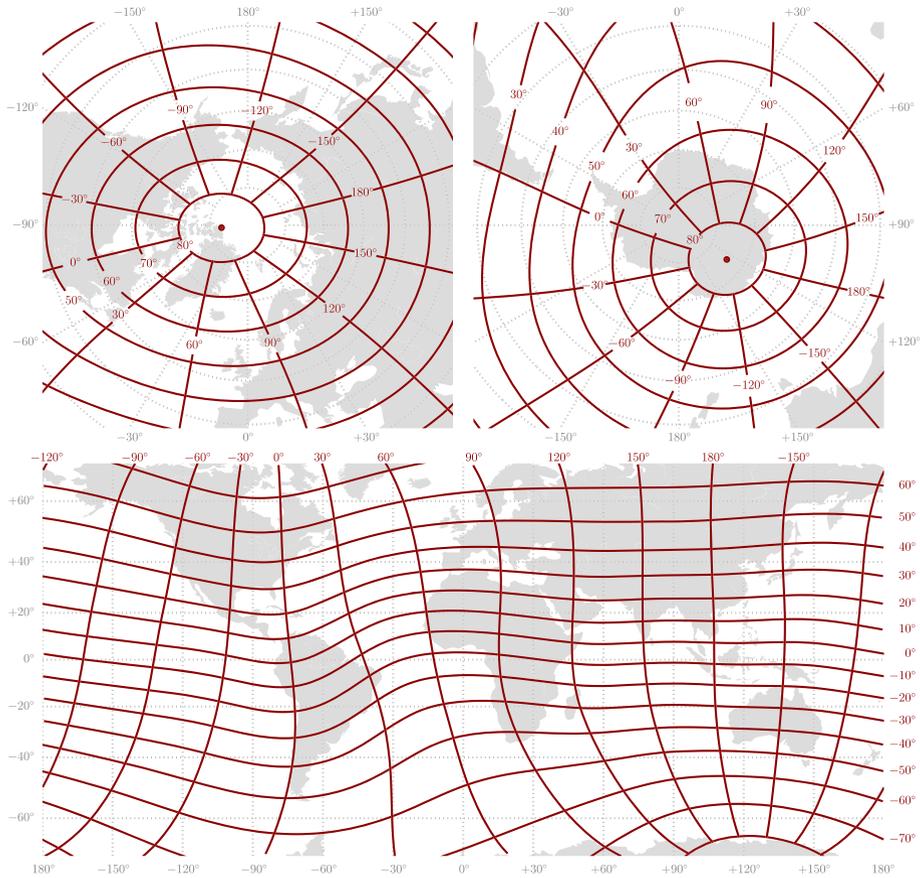

**Fig. 5** Maps of Quasi-Dipole coordinates at 0 km altitude, which are equal to Modified Apex coordinates with reference height 0. Calculated using code published by Emmert et al. (2010). The projections are similar to Fig. 1

to the IGRF magnetic field (10) as $\mathbf{B} = -\nabla V$. The third coordinate in Modified Apex coordinates is rarely explicitly used (in contrast to the third coordinate $h$ in QD coordinates), but it is useful for reasons that we will return to in Sect. 5.

Maps of Quasi-Dipole coordinates at $h = 0$ are shown in Fig. 5. These coordinates are the same as Modified Apex coordinates at $h = 0$ with $h_R = 0$. Notice that the grid is non-uniform. A global QD grid is also shown in Fig. 6 in a conformal projection, which means that local angles are preserved. It is clear that the coordinate contours in general do not intersect at right angles. The angle between eastward and northward unit vectors are indicated in color in Fig. 6, showing that the angles range from $\approx 60°$ to $\approx 116°$. The deviation from orthogonality is particularly significant in the South Atlantic and in the southern parts of Africa.

Direct calculation of apex coordinates by field line tracing is computationally expensive and complicated to implement. Emmert et al. (2010) published Fortran software for computationally compact conversion between apex and geodetic coordinates, using

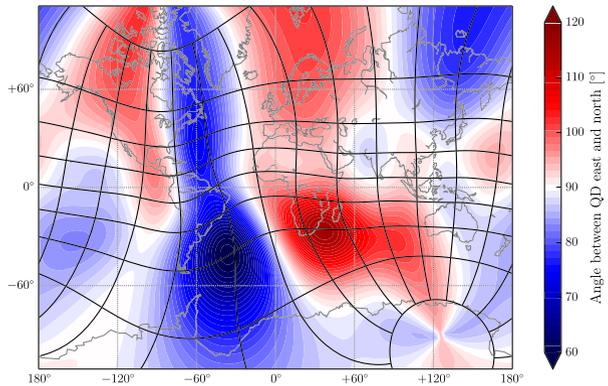

**Fig. 6** *Color* representation of the angle between eastward and northward unit vectors in the Quasi-Dipole coordinate system. A QD grid is overlaid. The map is shown in Mercator projection, which preserves angles. Computed using software by Emmert et al. (2010) with IGRF-12 2015 coefficients

spherical harmonic analysis. Their code has been used to produce the figures in this paper.[1]

Application of apex coordinates is only possible for planets like Earth whose magnetic field is not very distorted. On Mars for example, every point will not have unique MA or QD coordinates, since the same apex height (and consequently latitude), can appear multiple times along a given dipole longitude. This is because Mars lacks a dynamo in the core, so that small-scale magnetic patches in the crust dominate.

### 4.2 Corrected Geomagnetic Coordinates (CGM)

Corrected geomagnetic coordinates (CGM) are also defined in terms of field line tracing using the IGRF. The CGM coordinates of a point $P$ are determined by the location of the intersection of the field line at $P$ with the centered dipole equatorial plane. A dipole mapping to 1 $R_E$ is applied to find the latitude:

$$\lambda_{cgm} = \pm \cos^{-1} \sqrt{\frac{R_E}{R_E + h_{eq}}}, \qquad (44)$$

where $h_{eq}$ is the height of the CD equatorial plane intersection. The sign of the latitude is negative when mapping to the SH and positive when mapping to the NH. The CGM longitude is the longitude of the intersection. Since the coordinates are completely determined by a certain point on the coinciding IGRF field line, each point on that field line will have the same coordinates.

It is clear from the definition in (44) that CGM coordinates will be very similar to Modified Apex coordinates with $h_R = 0$. The difference between these coordinates depends on the difference between the location of the field line apex and the intersection with the CD equatorial plane. Figure 7 shows a CGM coordinate grid at $h = 0$, plotted on top of the QD grid (or MA with $h_R = 0$) from Fig. 5(blue). It can be seen that on these scales, there are only very small differences at polar latitudes. This is not surprising, since at high latitudes the apexes of the field lines are at very large altitudes where the dipole terms in the IGRF dominate. Consequently the apexes will be close to the dipole equatorial plane.

---

[1] A Python wrapper for this code is available at https://github.com/cmeeren/apexpy (by Christer van der Meeren and Karl M. Laundal, University in Bergen).

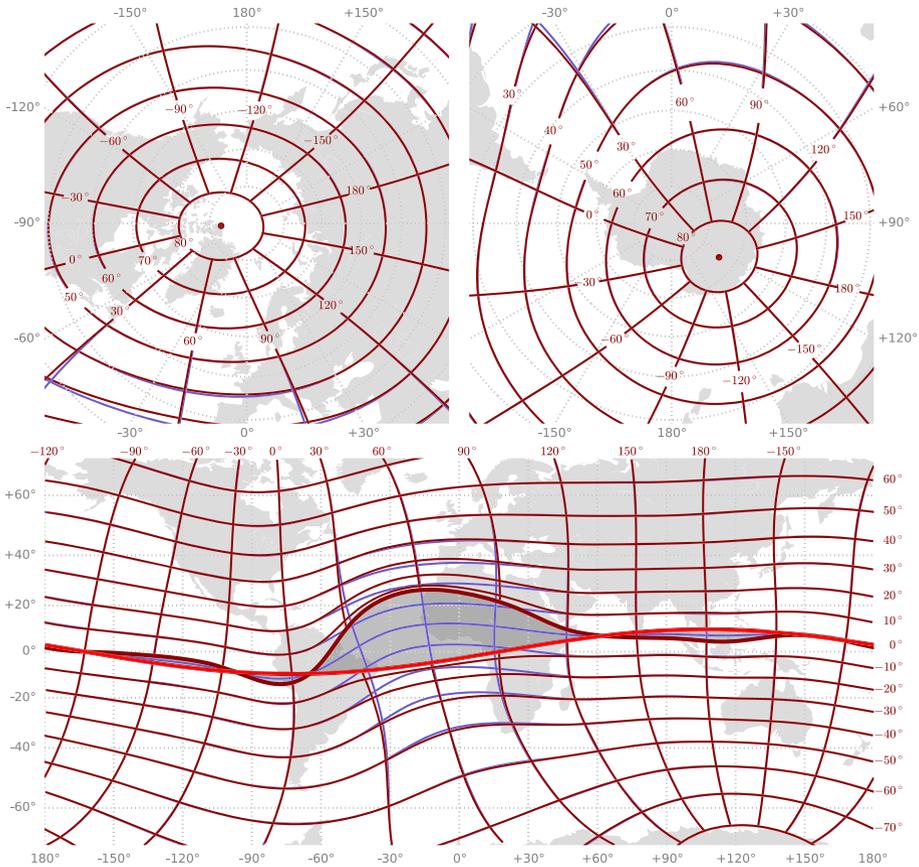

**Fig. 7** CGM (or AACGM) coordinate grid at zero height (*red*) plotted on top of the QD grid from Fig. 5 (*blue*). The CGM equator has been traced to both hemispheres (using the software by Shepherd 2014), and is shown as *two bold contours*. CGM coordinates are undefined in the *shaded regions* between these contours. The centered dipole equator is indicated in *bright red*. All conversions were done with IGRF12 2015 coefficients. The projections are similar to Fig. 1

At lower latitudes however the differences can be significant. In some regions surrounding the QD equator (which is also the dip equator), the CGM coordinates are not even defined, since there are field lines which never cross the CD equatorial plane. The CGM equator ($\lambda_{cgm} = 0$) has been mapped to both hemispheres, and is shown as bold contours in Fig. 7. Between these contours, in the shaded areas, CGM coordinates are undefined. The CD equator, marked with bright red, lies at one of the boundaries and the other is traced by points which are magnetically conjugate (i.e., at the other field line intersection with $h = 0$) to the CD equator. The CD equator is far south of its conjugate points in the Atlantic sector, but the orientation is reversed at some points along the dip equator.

Early implementations of CGM coordinate conversions were based on lookup tables (see e.g., Gustafsson et al. 1992, and references therein), and conversions were only computed for points at $h = 0$. Thus CGM coordinates are sometimes described as undefined at $h > 0$. Later implementations, using spherical harmonics (Baker and Wing 1989), allowed conversion of points at higher altitudes, and has become known as altitude-adjusted corrected geomagnetic (AACGM) coordinates. The definition of AACGM coordinates is mathematically the same as that described above for CGM coordinates. Despite the CGM/AACGM coordinates being undefined in certain regions, computer implementations prior to Shepherd (2014) have provided coordinates in these regions, using interpolation techniques. However Shepherd (2014) found that the interpolation introduced errors affecting also poleward latitudes. His code is more accurate than earlier software in the region where the coordinates are defined. The CGM coordinates in Fig. 7 were calculated by field line tracing using the software published by Shepherd (2014), which is available in the C and IDL languages.[2] The software also contains code to calculate conversions using spherical harmonics, which is much faster than with field line tracing.

## 5 Vector Quantities in Nonorthogonal Coordinates

The fact that CGM and apex coordinates are nonorthogonal implies that conversion of vectors is more complicated than a multiplication by a rotation matrix. In this section we describe how to do such conversions, starting by briefly reviewing a text-book description on nonorthogonal coordinates. Then we describe the technique introduced by Richmond (1995), which leads to certain very convenient properties for electrodynamic quantities regarding mapping along magnetic field lines. This section is partly motivated by the fact that several authors use CGM or apex grids with vectors represented in e.g. CD or local magnetic coordinates. This practice is mathematically unsound, since the vectors, which should be invariant with respect to the coordinate system, are effectively changed. This can lead to systematic errors between longitudes and hemispheres (Gasda and Richmond 1998; Laundal and Gjerloev 2014). In Sect. 5.2 we look at such errors in a quantitative fashion.

Following the description in Riley et al. (2006, Chap. 21) (see also the book by D'haeseleer et al. 1991), we have that in a general three-dimensional curvilinear coordinate system, with coordinates $u_1, u_2, u_3$, there exists two sets of base vectors at a point $P$:

$$\boldsymbol{\epsilon}^i = \nabla u_i \tag{45}$$

$$\boldsymbol{\epsilon}_i = \frac{\partial \mathbf{r}}{\partial u_i} \tag{46}$$

where $i = 1, 2, 3$ and $\mathbf{r} = \mathbf{r}(u_1, u_2, u_3)$ is the position vector of $P$. These base vectors are reciprocal, meaning that

$$\boldsymbol{\epsilon}_i \cdot \boldsymbol{\epsilon}^j = \delta_{ij}. \tag{47}$$

This property means that any vector $\mathbf{v}$ can be expressed in terms of either of the bases by:

$$\mathbf{v} = \sum_{i=1}^{3} (\boldsymbol{\epsilon}^i \cdot \mathbf{v}) \boldsymbol{\epsilon}_i = \sum_{i=1}^{3} (\boldsymbol{\epsilon}_i \cdot \mathbf{v}) \boldsymbol{\epsilon}^i. \tag{48}$$

The vectors $\boldsymbol{\epsilon}_i$ are called the covariant base vectors, and $\boldsymbol{\epsilon}^i$ contravariant base vectors. Similarly, the components $\boldsymbol{\epsilon}^i \cdot \mathbf{v} = v^i$ ($\boldsymbol{\epsilon}_i \cdot \mathbf{v} = v_i$) are called the contravariant (covariant) components of $\mathbf{v}$. Either decomposition is valid, since both keep the vector invariant with respect to the coordinate systems.

---

[2]A Python wrapper for this code is available at https://github.com/cmeeren/aacgmv2 (by Christer van der Meeren and Karl M. Laundal, University in Bergen).

This is a generalization of base vectors in Cartesian coordinate systems, where $\boldsymbol{\epsilon}^i = \boldsymbol{\epsilon}_i$, and the base vectors are orthogonal and of unit length. For example, the spherical coordinate base vectors $\hat{\mathbf{e}}, -\hat{\mathbf{n}}, \hat{\mathbf{u}}$ (4) can be derived by calculating $\frac{\partial \mathbf{r}}{\partial u_i}/\|\frac{\partial \mathbf{r}}{\partial u_i}\|$, with $\mathbf{r}$ given on the right hand side of (1), and $u_i = r, \theta, \phi$. The same result would appear by calculating the gradient of the spherical coordinates (45). Notice the change in sign of $\hat{\mathbf{n}}$, since $\frac{\partial \mathbf{r}}{\partial \theta}$ is southward when $\theta$ is colatitude. If $\mathbf{r}$ was differentiated with respect to the latitude, the result would be northward.

In a non-Cartesian coordinate system, such as CGM and Magnetic Apex Coordinates, $\boldsymbol{\epsilon}^i$ is in general not equal to $\boldsymbol{\epsilon}_i$. $\boldsymbol{\epsilon}^i$ is perpendicular to surfaces of constant $u_i$, and its length depends on the spacing between these surfaces. $\boldsymbol{\epsilon}_i$ is tangent to the contours defined by the intersection between the other two coordinate surfaces, and its length depends on the rate of change of $u_i$ along these contours.

This formalism can also be used to derive expressions for differential operators in non-Cartesian coordinate systems. The gradient, divergence, and curl operations for a right-handed system can be written

$$\nabla \phi = \sum_{i=1}^{3} \frac{\partial \phi}{\partial u_i} \boldsymbol{\epsilon}^i. \tag{49}$$

$$\nabla \cdot \mathbf{J} = \frac{1}{W} \sum_{i=1}^{3} \frac{\partial (W \boldsymbol{\epsilon}^i \cdot \mathbf{J})}{\partial u_i} \tag{50}$$

$$\nabla \times \mathbf{B} = \sum_{i=1}^{3} \frac{\boldsymbol{\epsilon}_i}{W} \sum_{j=1}^{3} \frac{\partial (W \boldsymbol{\epsilon}^i \times \boldsymbol{\epsilon}^j \cdot \mathbf{B})}{\partial u_i} \tag{51}$$

$$W = \left( \boldsymbol{\epsilon}^1 \times \boldsymbol{\epsilon}^2 \cdot \boldsymbol{\epsilon}^3 \right)^{-1} \tag{52}$$

### 5.1 Apex Base Vectors

It is tempting to apply (45) and (46) directly to Modified Apex coordinates to obtain base vectors for this coordinate system. However, Richmond (1995) presented base vectors which are scaled such that if the Earth's field was a perfect dipole, they would constitute an orthonormal set of vectors on the sphere $r = R_E + h_R$. The appropriate scaling factors can be found by evaluating $\|\nabla u_i\|$, $u_i = (\lambda_{ma}, \phi_{ma}, V)$ for a dipole field. The resulting scaled vectors were called $\mathbf{d}_i$, and are defined as follows:

$$\mathbf{d}_1 = (R_E + h_R) \cos \lambda_{ma} \nabla \phi_{ma} \tag{53}$$

$$\mathbf{d}_2 = -(R_E + h_R) \sin I_{ma} \nabla \lambda_{ma} \tag{54}$$

$$\mathbf{d}_3 = \frac{-\nabla V}{|\nabla V| D} = \frac{\mathbf{d}_1 \times \mathbf{d}_2}{\|\mathbf{d}_1 \times \mathbf{d}_2\|^2} \tag{55}$$

where

$$\sin I_{ma} = 2 \sin \lambda_{ma} \left( 4 - 3 \cos^2 \lambda_{ma} \right)^{-1/2}. \tag{56}$$

In a dipole $\mathbf{d}_1$ would point horizontally eastward. $\mathbf{d}_2$ would point equatorward and downward, so that it is perpendicular to the inclined magnetic field. $\sin I_{ma}$ can be recognized as the equation for the magnetic field inclination $I_{ma}$ in a dipole field. If the field was a dipole

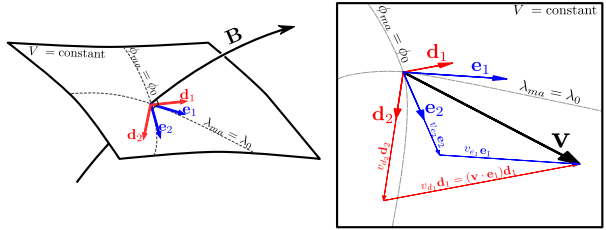

**Fig. 8** Illustration of the Modified Apex base vectors. *Left*: Base vectors with respect to **B**, $V$ and contours of constant coordinates. *Right*: Decomposition of a vector **v** in terms of $\mathbf{d}_i$ or $\mathbf{e}_i$ (shown on a surface of constant $V$)

and the Earth a sphere, $I_{ma}$ would be the inclination at radius $R_E + h_R$. $\mathbf{d}_3$ is field-aligned. The factor $D$, which would be unity in a dipole field at $r = R_E + h_R$ varies along field lines in proportion to the magnetic field strength $B$.

Since the field is not a dipole, $\mathbf{d}_1$ in general does not point strictly eastward along contours of constant Modified Apex latitude. It points perpendicular to Modified Apex meridians in surfaces of constant $V$. Likewise, $\mathbf{d}_2$ points roughly equatorward and downward, exactly perpendicular to contours of constant latitude in surfaces of constant $V$. Both $\mathbf{d}_1$ and $\mathbf{d}_2$ are perpendicular to the magnetic field. Because of the non-dipole terms, $\mathbf{d}_i$ will in general not have unit length at $r = R_E + h_R$, and the vectors will not be perpendicular to each other. The quantity $D = \|\mathbf{d}_1 \times \mathbf{d}_2\|$ can be seen as a measure of the deviation from a dipolar field at $r = R_E + h_R$. A map of $D$ can be found in the paper by Richmond (1995).

The second set of base vectors, $\mathbf{e}_i$, corresponding to $\boldsymbol{\epsilon}_i$ (46), are defined in terms of $\mathbf{d}_i$, so that the two sets are reciprocal ($\mathbf{d}_i \cdot \mathbf{e}_j = \delta_{ij}$):

$$\mathbf{e}_1 = \mathbf{d}_2 \times \mathbf{d}_3 \tag{57}$$

$$\mathbf{e}_2 = \mathbf{d}_3 \times \mathbf{d}_1 \tag{58}$$

$$\mathbf{e}_3 = \mathbf{d}_1 \times \mathbf{d}_2. \tag{59}$$

The $\mathbf{e}_1$ vector is tangent to coordinate contours defined by constant $\lambda_{ma}$ and constant $V$. Thus on a surface of constant $V$, $\mathbf{e}_1$ points eastward along contours of constant $\lambda_{ma}$, and $\mathbf{e}_2$ points equatorward along contours of constant $\phi_{ma}$ (magnetic meridians). $\mathbf{e}_3$ is aligned with the IGRF magnetic field, just as $\mathbf{d}_3$. The magnitude of $\mathbf{e}_3$ increases with altitude, while the magnitude of $\mathbf{d}_3$ decreases, such that the reciprocal property is fulfilled ($\mathbf{d}_3 \cdot \mathbf{e}_3 = 1$). $\mathbf{e}_1$ and $\mathbf{e}_2$ are in general not aligned with $\mathbf{d}_1$ and $\mathbf{d}_2$.

The left part of Fig. 8 shows an illustration of the base vectors with respect to **B**, $V$ and contours of constant coordinates. The right part of the figure, which displays a surface of constant $V$, shows how a vector **v** can be decomposed in terms of any of the bases, in the same way as in (48):

$$\mathbf{v} = v_{e_1}\mathbf{e}_1 + v_{e_2}\mathbf{e}_2 + v_{e_3}\mathbf{e}_3 = (\mathbf{v} \cdot \mathbf{d}_1)\mathbf{e}_1 + (\mathbf{v} \cdot \mathbf{d}_2)\mathbf{e}_2 + (\mathbf{v} \cdot \mathbf{d}_3)\mathbf{e}_3 \tag{60}$$

$$\mathbf{v} = v_{d_1}\mathbf{d}_1 + v_{d_2}\mathbf{d}_2 + v_{d_3}\mathbf{d}_3 = (\mathbf{v} \cdot \mathbf{e}_1)\mathbf{d}_1 + (\mathbf{v} \cdot \mathbf{e}_2)\mathbf{d}_2 + (\mathbf{v} \cdot \mathbf{e}_3)\mathbf{d}_3 \tag{61}$$

Which decomposition one should choose depends on the quantity (see Sect. 5.1.1).

Richmond (1995) also defined base vectors for Quasi-Dipole coordinates. For a dipole field and spherical Earth, both sets of QD base vectors would be equal to $\hat{\mathbf{e}}, \hat{\mathbf{n}}, \hat{\mathbf{u}}$ (4). Since the field is not a perfect dipole, they are generally nonorthogonal and not of unit length. They are defined as:

$$\mathbf{g}_1 = \frac{(R_E + h)}{F}\cos\lambda_{qd}\nabla\phi_{qd} \tag{62}$$

$$\mathbf{g}_2 = \frac{R_E + h}{F} \nabla \lambda_{qd} \tag{63}$$

$$\mathbf{g}_3 = F\mathbf{k} \tag{64}$$

and

$$\mathbf{f}_1 = \mathbf{g}_2 \times \mathbf{g}_3 \tag{65}$$

$$\mathbf{f}_2 = \mathbf{g}_3 \times \mathbf{g}_1 \tag{66}$$

$$\mathbf{f}_3 = \mathbf{g}_1 \times \mathbf{g}_2 \tag{67}$$

where $F = \|\mathbf{f}_1 \times \mathbf{f}_2\|$ and $\hat{\mathbf{k}}$ is an upward unit vector. $\mathbf{g}_i$ correspond to $\boldsymbol{\epsilon}^i$ and $\mathbf{f}_i$ to $\boldsymbol{\epsilon}_i$ according to

$$\boldsymbol{\epsilon}^1 = \frac{F}{(R_E + h)\cos\lambda_{qd}} \mathbf{g}_1 \tag{68}$$

$$\boldsymbol{\epsilon}^2 = \frac{F}{(R_E + h)} \mathbf{g}_2 \tag{69}$$

$$\boldsymbol{\epsilon}^3 = \frac{1}{F} \mathbf{g}_3 = \mathbf{k} \tag{70}$$

$$\boldsymbol{\epsilon}_1 = \frac{(R_E + h)\cos\lambda_{qd}}{F} \mathbf{f}_1 \tag{71}$$

$$\boldsymbol{\epsilon}_2 = \frac{(R_E + h)}{F} \mathbf{f}_2 \tag{72}$$

$$\boldsymbol{\epsilon}_3 = F\mathbf{f}_3 \tag{73}$$

and the volume factor $W$ of (52) is

$$W = \frac{(R_E + h)^2 \cos\lambda_{qd}}{F}. \tag{74}$$

$\mathbf{f}_1$ and $\mathbf{f}_2$ are horizontal vectors (on the surface of constant $h$) tangent to contours of constant $\lambda_{qd}$ and $\phi_{qd}$, respectively. The sign is such that $\mathbf{f}_1$ points eastward and $\mathbf{f}_2$ points northward.

As is clear from the definitions of the base vectors, the gradients of the magnetic coordinates must be calculated in order to find their components. Computer code which provides all base vectors (except for $\mathbf{g}_i$ and $\mathbf{f}_3$ which can easily be found when the others are known using relations in Richmond 1995) was published by Emmert et al. (2010).

### 5.1.1 Decomposing E, j, v and ∆B

Ionospheric electrodynamic quantities can in principle be decomposed in terms of either of the bases. The vector components will be different depending on which decomposition is chosen. In some cases a specific choice of decomposition leads to convenient mapping properties for the components. For example, if we consider the equation for the gradient in nonorthogonal coordinates (49), we see that it is decomposed in terms of $\nabla u_i$, which in the case of Modified Apex coordinates is parallel to $\mathbf{d}_i$. Thus, for the electric field, which in certain cases can be written as a gradient $\mathbf{E} = -\nabla \Phi$, it makes sense to decompose in terms of $\mathbf{d}_i$. Using the gradient in Modified Apex coordinates, $\mathbf{E}$ expressed in terms of $\Phi$ is

$$\mathbf{E} = (\mathbf{E} \cdot \mathbf{e}_1)\mathbf{d}_1 + (\mathbf{E} \cdot \mathbf{e}_2)\mathbf{d}_2$$

$$
\begin{aligned}
&= E_{d_1} \mathbf{d}_1 + E_{d_2} \mathbf{d}_2 \\
&= -\frac{1}{R_E + h_R} \left( \frac{1}{\cos \lambda_{ma}} \frac{\partial \Phi}{\partial \phi_m} \mathbf{d}_1 + \frac{1}{\sin I_{ma}} \frac{\partial \Phi}{\partial \lambda_m} \mathbf{d}_2 \right)
\end{aligned}
\quad (75)
$$

where the field-aligned component of $\mathbf{E}$ has been assumed to be zero. The equation shows that, since $\Phi$ maps along field lines, $E_{d_1}$ and $E_{d_2}$ must also be constant along field lines. The change in $\mathbf{E}$ along field lines is contained in the base vectors $\mathbf{d}_1$ and $\mathbf{d}_2$, which do change. This property can be used to map $\mathbf{E}$ along IGRF field lines. Because of the scaling of the vectors, the magnitude of the electric field components can be interpreted as the magnitude at $r = R_E + h_R$ in a dipole field.

If the main magnetic field (IGRF) $\mathbf{B}$ is decomposed in terms of $\mathbf{e}_3$,

$$
\mathbf{B} = (\mathbf{B} \cdot \mathbf{d}_3) \mathbf{e}_3 = B_{e_3} \mathbf{e}_3,
\quad (76)
$$

the quantity $B_{e_3}$ is constant along field lines.

These properties imply that the component of the drift velocity, $\mathbf{v} = \mathbf{E} \times \mathbf{B}/B^2$, can also be made constant along field lines if they are decomposed in terms of $\mathbf{e}_i$. That is,

$$
\mathbf{v} = \frac{E_{d_2}}{B_{e_3}} \mathbf{e}_1 - \frac{E_{d_1}}{B_{e_3}} \mathbf{e}_2 = (\mathbf{v} \cdot \mathbf{d}_1) \mathbf{e}_1 + (\mathbf{v} \cdot \mathbf{d}_2) \mathbf{e}_2 = v_{e_1} \mathbf{e}_1 + v_{e_2} \mathbf{e}_2,
\quad (77)
$$

where $v_{e_1}$ and $v_{e_2}$ are constant along field lines. Their magnitude can be interpreted as the drift magnitude at $r = R_E + h_R$ in a dipole field.

The electric current density $\mathbf{J}$ is also conveniently decomposed in terms of the $\mathbf{e}_i$ base vectors:

$$
\mathbf{J} = (\mathbf{J} \cdot \mathbf{d}_1) \mathbf{e}_1 + (\mathbf{J} \cdot \mathbf{d}_2) \mathbf{e}_2 + (\mathbf{J} \cdot \mathbf{d}_3) \mathbf{e}_3 = J_{e_1} \mathbf{e}_1 + J_{e_2} \mathbf{e}_2 + J_{e_3} \mathbf{e}_3.
\quad (78)
$$

Steady-state currents satisfy a force-balance condition

$$
\mathbf{J} \times \mathbf{B} + \mathbf{F}' = 0,
\quad (79)
$$

where $\mathbf{F}'$ is the force per unit volume on the charged particles due to collisions with neutral molecules, to gravity, and to plasma pressure gradients. If $\mathbf{F}'$ is also expressed in terms of the $\mathbf{e}_i$ base vectors, then (79) has the solution

$$
J_{e_1} = \frac{F'_{e_2}}{B_{e_3}}
\quad (80)
$$

$$
J_{e_2} = -\frac{F'_{e_1}}{B_{e_3}}.
\quad (81)
$$

When the current is force-free, as is approximately the case in the top-side ionosphere and lower magnetosphere, the current components $J_{e_1}$ and $J_{e_2}$ essentially vanish and $J_{e_3}$ is nearly constant along geomagnetic-field lines, such that the field-aligned current density varies along the field lines as $\mathbf{e}_3$.

In the lower ionosphere the constraints that current be continuous and that the vertical current essentially vanish at the bottom of the ionosphere result in the current being predominantly horizontal. It is therefore also useful to express $\mathbf{J}$ in terms of scaled QD components:

$$
\mathbf{J} = J_{f_1} \mathbf{f}_1 + J_{f_2} \mathbf{f}_2 + J_{f_3} \mathbf{f}_3 = (\mathbf{g}_1 \cdot \mathbf{J}) \mathbf{f}_1 + (\mathbf{g}_2 \cdot \mathbf{J}) \mathbf{f}_2 + (\mathbf{g}_3 \cdot \mathbf{J}) \mathbf{f}_3.
\quad (82)
$$

The first two components are horizontal, while the third component is approximately, but not precisely, vertical. That is, a small part of the current $J_{f_3}\mathbf{f}_3$ projects onto the horizontal plane, since $\mathbf{f}_3$ is not precisely vertical, in general. However, the value of $J_{f_3}$ is determined solely by the vertical current density, multiplied by the scaling factor $F$, since $\mathbf{g}_3$ is precisely vertical.

In magnetostatics the current density is proportional to the curl of the magnetic field. Since the main field is curl-free, $\mathbf{J}$ can be related to the curl of the magnetic-perturbation field $\Delta\mathbf{B}$:

$$\mathbf{J} = \nabla \times (\Delta\mathbf{B})/\mu_0 \qquad (83)$$

where $\mu_0$ is the permeability of free space. Applying (49), (51), (68)–(73), (74), and (83) we find that

$$J_{f_1} = \frac{1}{\mu_0(R_E+h)}\left[\frac{\partial \Delta B_{qr}}{\partial \lambda_{qd}} - \frac{\partial [(R_E+h)\Delta B_{q\lambda}]}{\partial h}\right] \qquad (84)$$

$$J_{f_2} = \frac{1}{\mu_0(R_E+h)}\left[\frac{\partial [(R_E+h)\Delta B_{q\phi}]}{\partial h} - \frac{1}{\cos\lambda_{qd}}\frac{\partial \Delta B_{qr}}{\partial \phi_{qd}}\right] \qquad (85)$$

$$J_{f_3} = \frac{F^2}{\mu_0(R_E+h)\cos\lambda_{qd}}\left[\frac{\partial \Delta B_{q\lambda}}{\partial \phi_{qd}} - \frac{\partial (\cos\lambda_{qd}\Delta B_{q\phi})}{\partial \lambda_{qd}}\right] \qquad (86)$$

$$\Delta B_{q\phi} = \frac{\mathbf{f}_1 \cdot \Delta\mathbf{B}}{F} \qquad (87)$$

$$\Delta B_{q\lambda} = \frac{\mathbf{f}_2 \cdot \Delta\mathbf{B}}{F} \qquad (88)$$

$$\Delta B_{qr} = F\mathbf{f}_3 \cdot \Delta\mathbf{B} \qquad (89)$$

In terms of the quantities (87)–(89) $\Delta\mathbf{B}$ is

$$\Delta\mathbf{B} = \Delta B_{q\phi} F\mathbf{g}_1 + \Delta B_{q\lambda} F\mathbf{g}_2 + \Delta B_{qr}\hat{\mathbf{k}}. \qquad (90)$$

Note that $\Delta B_{qr}$ is approximately, but not precisely, the vertical component of $\Delta\mathbf{B}$, because $\mathbf{g}_1$ and $\mathbf{g}_2$ have small vertical components, in general.

Although (84)–(89) can readily be used to calculate $\mathbf{J}$ from $\Delta\mathbf{B}$ in QD coordinates, the inverse calculation to get $\Delta\mathbf{B}$ from $\mathbf{J}$ would be considerably more difficult in QD coordinates than in geocentric spherical coordinates, because the Biot–Savart relation, or a spherical-harmonic representation of it, becomes very complicated in QD coordinates, owing to the complicated calculation of distances and the non-orthonormality of the base vectors.

Synoptic analyses of satellite or ground magnetic-perturbation data are often organized with respect to magnetic coordinates. Laundal and Gjerloev (2014) showed that calculation of the geomagnetic *SML* index (Newell and Gjerloev 2011) yields less longitude-dependent variation when computed with $\Delta B_{q\lambda}$ than when computed with the unscaled magnetic perturbations in the local magnetic-north direction. If satellite magnetic-perturbation data are binned with respect to magnetic latitude and magnetic local time, it can also be convenient to orient and scale them using the components $\Delta B_{q\phi}$, $\Delta B_{q\lambda}$, and $\Delta B_{qr}$ as defined in (87)–(89). One reason for this is that scaled current-density components can be calculated from (84)–(86) directly in QD coordinates without the need for longitude-dependent scale factors. $J_{f_1}$ and $J_{f_2}$, as calculated from (84) and (85), represent the scaled horizontal current density components. For the vertical current density, on the other hand, $J_{f_3}$ as computed from (86) does not itself make a convenient representation, because it depends not only on the QD coordinates, but also on the geographically varying scale factor $F^2$. Instead, the appropriate scaled vertical current density component is a quantity we can call $J_{qr}$, defined by

$$J_{qr} = J_{f_3}/F^2 = J_r/F. \tag{91}$$

$J_{qr}$ equals the quantity $J_{mr}$ defined by Richmond (1995) at the height of the reference altitude for Modified Magnetic Apex coordinates.

While the calculation of the divergence of $\mathbf{J}$ as expressed by (82) in QD coordinates is fairly straightforward (and results in identically zero divergence), it is not straightforward to calculate the divergence of $\Delta\mathbf{B}$ as expressed by (90) in QD coordinates. (One might wish to calculate the divergence of an empirical model of $\Delta\mathbf{B}$, for example, to ensure it is zero.) This difference in ease of calculation of the divergence is due to the fact that $\mathbf{J}$ in (82) is expressed in components related to covariant base vectors, while $\Delta\mathbf{B}$ in (90) is expressed in components related to contravariant base vectors.

Richmond (1995) presented a way to relate magnetic perturbations to three-dimensional currents partially using Modified Magnetic Apex coordinates, but requiring an additional calculation of a three-dimensional magnetic potential by solving Poisson's equation in geographic coordinates. That paper pointed out how the source term for the Poisson's equation can be made relatively small by first solving for a suitable two-dimensional function, constant along field lines, that minimizes the source term in a region of interest. Apart from the component of the magnetic perturbation associated with the solution of the Poisson's equation, the remaining portion of magnetic perturbations associated with field-aligned currents maps along geomagnetic-field lines. ($\Delta B_{q\phi}$, $\Delta B_{q\lambda}$, and $\Delta B_{qr}$ do not map along field lines.) We are not aware of any attempt to implement the procedure outlined in that paper with numerical algorithms.

### 5.2 Applications/Examples

Modified Magnetic Apex and Quasi-Dipole coordinates can be useful for examining how plasma velocities vary along a geomagnetic field line in the ionosphere.

An example is shown in Fig. 9. The left half of this figure answers the following question: if we look at a point in the ionosphere at 250 km altitude at some location on the Earth, and if the $\mathbf{E} \times \mathbf{B}/B^2$ velocity at the conjugate point in the opposite hemisphere at 250 km altitude has a unit value in the magnetic-eastward direction $\hat{\mathbf{e}}_m^{conj}$ (as defined in Sect. 3.2), then what are the local velocity components in the magnetic-eastward $\hat{\mathbf{e}}_m$ and magnetic-meridional $\hat{\mathbf{p}}$ directions? We calculate this by setting $\mathbf{v}^{conj} = \hat{\mathbf{e}}_m^{conj}$, computing $v_{e1}$ and $v_{e2}$ by taking the scalar products of $\mathbf{v}^{conj}$ with $\mathbf{d}_1^{conj}$ and $\mathbf{d}_2^{conj}$, mapping $v_{e1}$ and $v_{e2}$ along the magnetic field to the local point ($v_{e1}$ and $v_{e2}$ being constant along field lines), computing the local velocity as $\mathbf{v} = v_{e1}\mathbf{e}_1 + v_{e2}\mathbf{e}_2$, and then computing the velocity components in the local magnetic-eastward and magnetic-meridional directions by taking the scalar product of $\mathbf{v}$ with the local orthogonal unit vectors $\hat{\mathbf{e}}_m$ and $\hat{\mathbf{p}}$. Figure 9(a) shows the resultant magnetic-eastward velocity, and Fig. 9(b) shows the resultant magnetic-meridional velocity. By a similar procedure we can start from a unit velocity in the magnetic-meridional direction $\hat{\mathbf{p}}_m^{conj}$ at the conjugate point, map it along the magnetic field to the local point, and calculate the local magnetic-eastward and magnetic-meridional components. These are shown in Fig. 9(c) and (d). For a purely dipolar geomagnetic field on a spherical Earth all the values in Fig. 9(a) and (d) would be 1, and all the values in Fig. 9(b) and (c) would be 0. In Fig. 9 values greater than these idealized dipolar values are plotted with red contours, while values less than the dipolar values are plotted with blue contours.

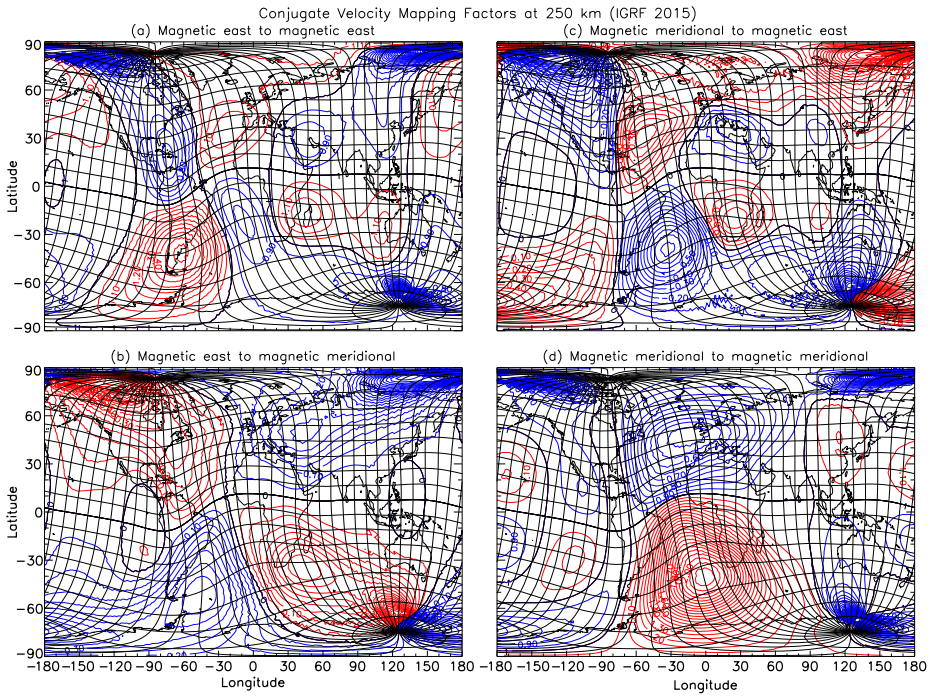

**Fig. 9** Conjugate velocity mapping factors at an altitude of 250 km for the IGRF epoch 2015.0. See the main text for explanation. The contour interval is 0.05

We can note some of the larger deviations from a purely dipolar mapping. A magnetic-eastward velocity of 1 m/s at 250 km altitude, 30°N QD latitude, corresponds at 30°S QD latitude to a magnetic-eastward velocity ranging from as small as 0.84 m/s in the middle South Atlantic Ocean to 1.51 m/s on the eastern coast of South America [Fig. 9(a)]. Owing to the twist of magnetic flux tubes, a purely horizontal magnetic-eastward velocity at 40°N QD latitude acquires a magnetic-meridional component that at 40°S QD latitude can be as large as 0.43 m/s downward/equatorward around −45°E geographic longitude, or 0.43 m/s upward/poleward around +45°E geographic longitude [Fig. 9(b)]. A magnetic-meridional velocity of 1 m/s at 250 km altitude, 44°N QD latitude, and 60°E QD longitude (the QD longitude line that grazes the coast of West Africa) corresponds at 44°S QD latitude to about 2.24 m/s in the local magnetic-meridional direction [Fig. 9(d)]. Above about 70° magnetic latitude the local magnetic-eastward and magnetic-meridional directions can deviate greatly from those of a tilted dipole, and the mapping factors can differ greatly from the dipole values of 0 or 1.

## 6 Magnetic Local Time

The CD, ED, CGM, QD, and MA coordinates are all fixed with respect to the Earth. It is often appropriate however to introduce a magnetic local time, instead of magnetic longitude, in order to organize data and models with respect to the position of the Sun (Vegard 1912, 1917).

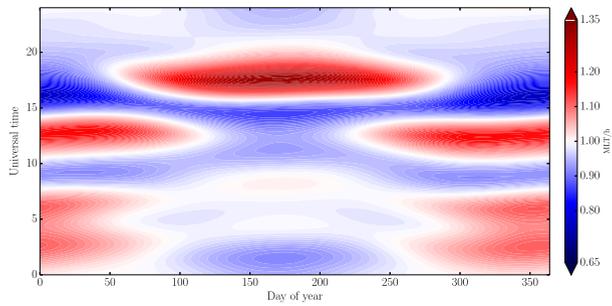

**Fig. 10** The rotation rate of the MLT/magnetic latitude grid with respect to an Earth-fixed grid. The apex longitude of the subsolar point (at height = 0) was computed with IGRF-12 2015 coefficients

A common definition of the magnetic local time is (e.g., Baker and Wing 1989):

$$\text{MLT} = \text{UT} + (\phi + \Phi_N)/15 \tag{92}$$

where $\phi$ is the magnetic longitude of the point in question, $\Phi_N$ is the geographic longitude of the North CD pole (14) and UT is the universal time specified in hours. Although simple to calculate, this definition does not generally yield MLT = 12 for the magnetic longitude of the subsolar point, as would Vegard's original definition applied to the dipole. Therefore, other definitions of MLT that take into consideration the magnetic longitude of the subsolar point are commonly used.

One such definition of MLT is the hour angle (1 hour is 15° magnetic longitude) from the midnight magnetic meridian, positive in the magnetic eastward direction. The midnight magnetic meridian can be defined as the meridian that is 180° magnetic longitude away from the subsolar point. The MLT/magnetic latitude coordinate system will then rotate with respect to the Earth at the rate at which the subsolar point crosses magnetic meridians.

As can be seen in Fig. 7, the spacing between magnetic meridians in CGM and apex coordinates is not constant. That implies that the above definition of the midnight meridian leads to a varying rotation rate of the MLT/magnetic latitude system with respect to the longitude/latitude system. The rotation rate will vary with the latitude of the track of the subsolar point, which depends on the season (see also Ono 1987). The rotation rate for a full year (2015) is shown in Fig. 10. The figure shows that the rotation rate can be more than 1.3 hours MLT per hour, and less than 0.8 hours MLT per hour.

The orientation of the dipole part of the Earth's field is more important for the Sun/Earth interaction than the orientation of the low-altitude non-dipole field structures. A recommended definition of MLT is therefore

$$\text{MLT} = (\phi - \phi_{cd,\hat{s}})/15 + 12, \tag{93}$$

where $\phi_{cd,\hat{s}}$ is the CD longitude of the subsolar point, and the local magnetic longitude $\phi$ can be in CD, ED, CGM/AACGM, or QD/MA coordinates, as long as it is made clear which is used. The CD longitude of the subsolar point, $R_E\hat{s}$, is similar to the QD/MA and CGM/AACGM longitude of $r\hat{s}$, with $r \gg 1R_E$. With this definition, the rotation rate in Fig. 10 reduces to the range 0.94–1.10 hours MLT per hour (for 2015).

To illustrate the effect that the different definitions of MLT may have on scientific results, and the importance of specifying which definition is used, we look at the magnetic local times of substorm onsets observed by the Imager for Magnetopause-to-Aurora Global Exploration (IMAGE) Far Ultraviolet Imager (FUV) cameras and by the Polar Ultraviolet Imager (UVI) camera. Lists of onset locations from these two satellites have been published

Magnetic Coordinate Systems

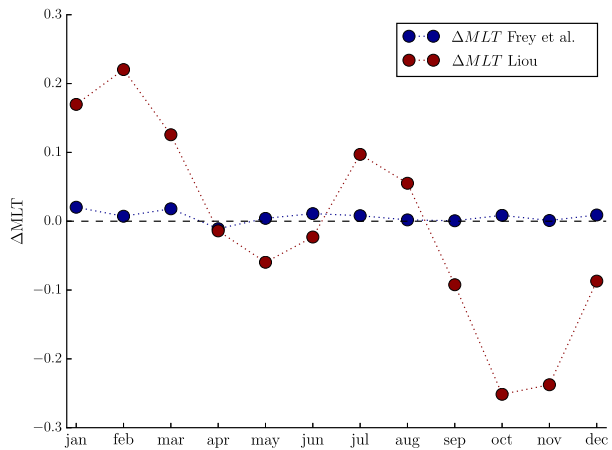

**Fig. 11** The monthly median difference between the substorm onset magnetic local time calculated by us and by Liou (2010) (*red*) and Frey et al. (2004) (*blue*). The figure shows that there is a systematic difference in how MLT is calculated in the two papers

by Frey et al. (2004) and Liou (2010), respectively. The lists have been extensively used for studies of e.g., the influence of the interplanetary magnetic field and dipole tilt angle on the magnetosphere (e.g., Østgaard et al. 2011, and references therein). We have calculated the magnetic local time of all the onset locations, using Modified Apex coordinates with $h_R = 130$ km, based on the geographic coordinates reported in the lists. We then calculated the monthly median difference between our calculations and those reported by Frey et al. (2004) and Liou (2010). The results are shown in Fig. 11. We see that our results largely match those by Frey et al. (2004), but that there is a significant and systematic difference with the MLTs reported by Liou (2010). That is not to say that one is better than the other (neither provide details on how the calculations of MLT were done), only that the calculations of MLT are done differently. The magnitude of the difference is modest, only up to 0.25 h, which rarely matters in studies of single events. However, since the difference is systematic, this may lead to wrong conclusions about the relation of substorm onset location to other parameters. For instance, Liou and Newell (2010) report that the average MLT of substorm onsets changes by 0.1 h for every 10° of tilt angle in the NH. This is the same order of magnitude as the differences seen in Fig. 11, and consequently the result might have been different if the MLT was calculated in the same way as in Frey et al. (2004).

## 7 Secular Variations

Each magnetic coordinate system is associated with two magnetic poles. Trajectories for four differently defined magnetic poles, the CD, ED, dip,[3] and QD/MA poles, are shown in Fig. 12. Their positions in 1950 are marked with large circles. Later positions are marked at 5 years intervals, ending in 2015. The centered dipole poles mark the intersection of the centered dipole axis with the Earth's surface. For all the other poles, the axes connecting them will be eccentric with respect to the center of the Earth. We see that the eccentric dipole poles and apex poles do not coincide at the surface. At very high altitudes, all the poles approach the centered dipole poles.

---

[3]The dip pole locations were obtained from http://www.geomag.bgs.ac.uk/education/poles.html.

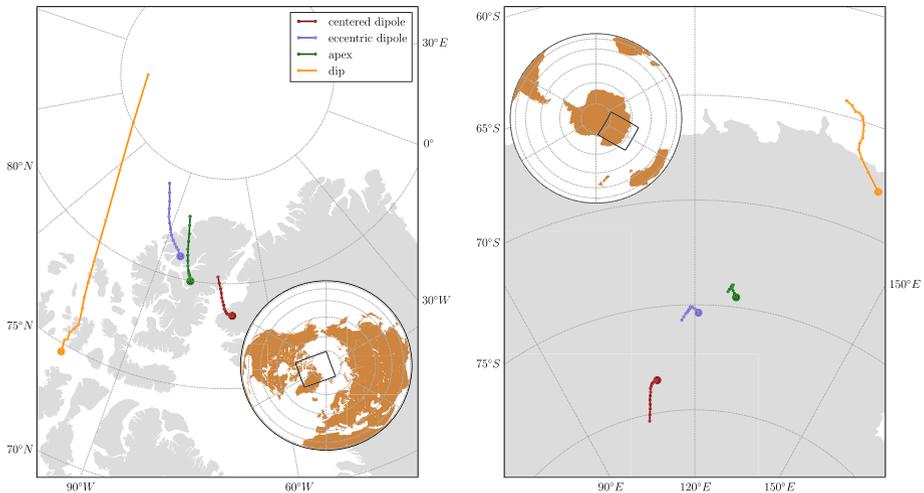

**Fig. 12** Positions of four differently defined magnetic poles. The positions in 1950 are marked by *large circles*. *Smaller circles* mark positions at steps of five years, ending in 2015. Based on IGRF-12. Lambert conformal conic projection is used in these plots

To further illustrate the time dependence of the magnetic coordinates we show in Fig. 13 QD grids for both 2015 (red) and 1985 (blue). We see that during the 30 year period since 1985, the changes in the field have led to significant changes in the QD coordinate grid. The entire grid has shifted northward nonuniformly, particularly in the region at 30° QD longitude. The magnitude of the change on this meridian is approximately 2° latitude at Northern polar latitudes and 5° latitude at mid and low latitudes.

## 8 Closing Remarks

We have presented the definitions of eight different magnetic coordinate systems: The centered dipole, eccentric dipole, solar magnetic, geocentric solar magnetic, corrected geomagnetic, Quasi-Dipole, and Modified Apex coordinates, as well as the dip latitude. This is not a complete list of magnetic coordinate systems that are or have been in use. In particular, the $B, L$ coordinate system, defined by McIlwain (1961, 1966) is much used in studies of trapped energetic particles in the inner magnetosphere. In that system $B$ is the magnetic field strength, and $L$ is a parameter which depends on $B$ and the longitudinal adiabatic invariant, $I$. $L$ is defined using a realistic field model, but in a dipole field $L$ would correspond to the radial distance, in $R_E$, of the field line. Thus O'Brien et al. (1962) introduced the invariant latitude $\Lambda$, which is based on $L$, using an equation which is similar to that which defines CGM coordinates:

$$\Lambda = \cos^{-1}\sqrt{\frac{1}{L}}, \tag{94}$$

which is valid in regions where $L \geq 1$. There are regions at low latitudes where $\Lambda$ is undefined.

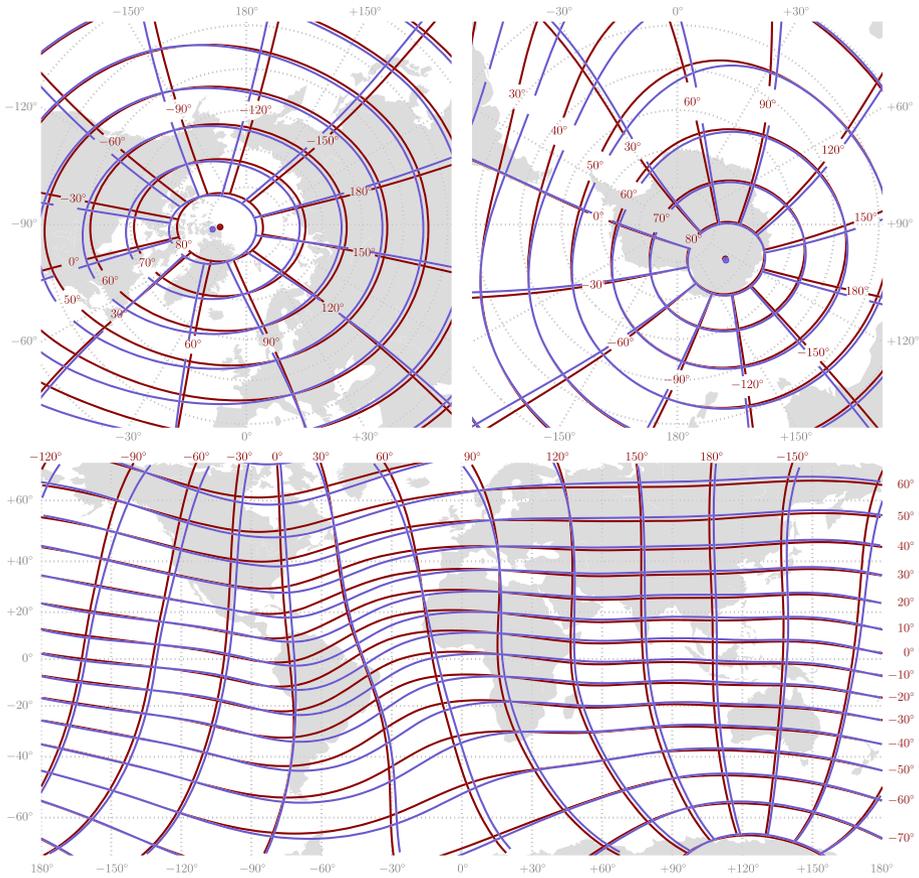

**Fig. 13** QD coordinate grids for epochs 2015 (*red*) and 1985 (*blue*). The projections are similar to Fig. 1

Another magnetic coordinate system, the magnetospheric geomagnetic latitude, was defined by Papitashvili et al. (1997a,b). This system takes into account also the external, highly time varying magnetospheric field by means of the Tsyganenko (1989) model.

The terminology associated with magnetic coordinates can be confusing and sometimes ambiguous. Traditionally, the term "geomagnetic" has been used to signify the poles, $z$ axis, and coordinate system associated with what we have called the centered dipole coordinate system (Vestine 1967; Chapman 1963). The term "magnetic" is often used to refer to the dip latitude, dip equator and dip poles, although it can also signify other systems. The terminology used in this paper is chosen because it is well established, and/or descriptive in terms of the definition of the coordinate system.

With at least six differently defined "magnetic latitudes", and several different definitions of MLT, it is clearly important to be precise about which coordinate system is used, and how MLT was calculated. Since the coordinate systems change in time, the epoch of the geomagnetic field used in the coordinate conversions should also be specified.

## Appendix A: Abbreviations

The following abbreviations are used to denote the various coordinate systems. They are used in uppercase in the text, and lowercase as subscripts in equations. Section number in parentheses.

| | |
|---|---|
| AACGM | Altitude-Adjusted Corrected Geomagnetic Coordinates (Sect. 4.2) |
| CD | Centered Dipole (Sect. 3.1) |
| CGM | Corrected Geomagnetic Coordinates (Sect. 4.2) |
| DIP | Dip (latitude) (Sect. 3.2) |
| ECEF | Earth-Centered Earth-Fixed (Sect. 2) |
| ED | Eccentric Dipole (Sect. 3.3) |
| ENU | East North Up (Sect. 2) |
| GEO | Geocentric (Sect. 2) |
| GSE | Geocentric Solar Ecliptic (Sect. 3.4) |
| GSM | Geocentric Solar Magnetic (Sect. 3.4) |
| M | Local Magnetic Coordinates (Sect. 3.2) |
| MA | Modified Apex (Sect. 4.1) |
| QD | Quasi-Dipole (Sect. 4.1) |
| SM | Solar Magnetic (Sect. 3.5) |

The following abbreviations are also used repeatedly throughout the paper

| | |
|---|---|
| IGRF | International Geomagnetic Reference Field |
| MLT | Magnetic Local Time |
| NH | Northern Hemisphere |
| SH | Southern Hemisphere |
| UT | Universal Time |
| WGS84 | World Geodetic System 1984 |

## Appendix B: Notation

The following notations is used for coordinates (with subscripts denoting the coordinate system):

| | |
|---|---|
| $x, y, z$ | Cartesian coordinates |
| $\theta$ | Polar angle (colatitude) |
| $\phi$ | Longitude (positive east) |
| $r$ | Radius |
| $\lambda$ | Latitude |
| $h$ | Geodetic height |
| $L$ | McIlwain's $L$ parameter |
| $\Lambda$ | Invariant latitude associated with $L$ |

The following symbols are also used repeatedly:

| | |
|---|---|
| $\hat{\mathbf{x}}$ | Unit vector in $x$-direction |
| $\hat{\mathbf{y}}$ | Unit vector in $y$-direction |
| $\hat{\mathbf{z}}$ | Unit vector in $z$-direction |
| $\hat{\mathbf{e}}$ | Unit vector in east-direction |
| $\hat{\mathbf{n}}$ | Unit vector in north-direction |
| $\hat{\mathbf{u}}$ | Unit vector in up/radial direction (geocentric) |
| $\hat{\mathbf{k}}$ | Unit vector in up-direction (geodetic) |
| $\hat{\mathbf{m}}$ | Unit vector in negative centered dipole axis direction (positive to the north) |
| $\hat{\mathbf{s}}$ | Unit vector pointing at subsolar point |
| $\hat{\mathbf{b}}$ | Unit vector in the direction of the magnetic field |
| $\mathbf{B}$ | Magnetic field vector |
| $g_n^m, h_n^m$ | Gauss coefficients in the IGRF |
| $R$ | Reference Earth radius used in geomagnetic modeling (6371.2 km) |
| $R_E$ | Mean Earth radius ($\approx$ 6371.009 km in WGS84) |
| $R_{eq}$ | Equatorial radius of the ellipsoid (6378.1370 km in WGS84) |
| $R_p$ | Polar radius of the ellipsoid (6356.7523 km in WGS84) |
| $h_R$ | Modified apex reference height |
| $h_A$ | Field line apex height |
| $h_{eq}$ | Height at which field line crosses dipole equatorial plane |
| $\psi$ | Dipole tilt angle |
| $H$ | Horizontal magnetic field component |
| $Z$ | Downward magnetic field component |
| $D$ | Magnetic declination angle |
| $I$ | Dip angle |
| $\mathbf{d}_i$ | Modified Apex base vector ($i = 1, 2, 3$) |
| $\mathbf{e}_i$ | Modified Apex base vector ($i = 1, 2, 3$) |
| $\mathbf{f}_i$ | Quasi-Dipole base vector ($i = 1, 2, 3$) |
| $\mathbf{g}_i$ | Quasi-Dipole base vector ($i = 1, 2, 3$) |
| $\boldsymbol{\epsilon}^i$ | Contravariant base vector |
| $\boldsymbol{\epsilon}_i$ | Covariant base vector |

## Appendix C: The Subsolar Point

The code below can be used to calculate the subsolar point, the location on Earth where the Sun is in zenith. The output is the co-latitude (`sbslcolat`) and longitude (`sbsllon`) of the subsolar point, and the inputs are `year` (integer, 4 digits), `doy` (day of year, integer, with `doy` being 1 on 1 January) and `ut` (universal time in seconds, possibly including a fraction, since the start of the day). The code can be run in Python, with the following import statement:

```
from math import floor, sin, cos, pi, atan2, asin
```

The code is based on formulas in Astronomical Almanac for the year 1996, p. C24.(U.S. Government Printing Office, 1994). It is usable for the years 1601–2100, inclusive. According to the Almanac, results are good to at least 0.01 degree latitude and 0.025 degree longitude between years 1950 and 2050. Accuracy for other years has not been tested. Every day is assumed to have exactly 86400 seconds; thus leap seconds that sometimes occur on December 31 are ignored. Their effect is below the accuracy threshold of the algorithm.

```
yr = year − 2000
nleap = floor((year−1601)/4.)
nleap = nleap − 99
if year <= 1900:
    ncent = floor((year−1601)/100.)
    ncent = 3 − ncent
    nleap = nleap + ncent
l0 = −79.549+(−.238699*(yr−4*nleap)+ 3.08514e−2*nleap)
g0 = −2.472+(−.2558905*(yr−4*nleap)− 3.79617e−2*nleap)
df = (ut/86400. − 1.5)+doy
lf = .9856474*df
gf = .9856003*df
l = l0 + lf
g = g0 + gf
grad = g*pi/180.
lmbda = l + 1.915*sin(grad) + .020*sin(2.*grad)
lmrad = lmbda*pi/180.
sinlm = sin(lmrad)
n = df + 365.*yr + nleap
epsilon = 23.439 − 4.0e−7*n
epsrad = epsilon*pi/180.
alpha = atan2(cos(epsrad)*sinlm, cos(lmrad)) * 180./pi
delta = asin(sin(epsrad)*sinlm) * 180./pi
sbslcolat = 90 − delta
etdeg = l − alpha
nrot = round(etdeg/360.)
etdeg = etdeg − 360.*nrot
aptime = ut/240. + etdeg
sbsllon = 180. − aptime
nrot = round(sbsllon/360.)
sbsllon = sbsllon − 360.*nrot
```